\documentclass{IEEEtaes}

\usepackage[utf8]{inputenc}
\usepackage{textcomp}
\usepackage{changes}
\usepackage{cuted}
\usepackage{amsfonts}
\usepackage{hyperref}  
\setaddedmarkup{\textcolor{red}{#1}}
\setdeletedmarkup{\textcolor{red}{\sout{#1}}}
\usepackage{float}
\usepackage{amsthm}
\theoremstyle{definition}
\newtheorem{theorem}{Theorem}[section]

\theoremstyle{definition}
\newtheorem{definition}[theorem]{Definition}

\theoremstyle{remark}
\newtheorem*{remark}{Remark}

\usepackage{graphicx}
\usepackage{amsmath}
\usepackage{float}
\usepackage[version=4]{mhchem}
\usepackage{siunitx}
\usepackage{longtable,tabularx}
\usepackage{bm}
\usepackage{cite}
\jvol{XX}
\jnum{XX}
\jmonth{XXXXX}
\paper{1234567}
\pubyear{2026}
\doiinfo{TAES.2026.Doi Number}

\begin{document}

\title{Constructing Four-Body Ballistic Lunar Transfers via Analytical Energy Conditions}

\author{Shuyue Fu}
\affil{Beihang University, Beijing, People's Republic of China}
\author{Di Wu}
\affil{Tsinghua University, Beijing, People's Republic of China}
\author{Xiaowen Liu}
\author{Peng Shi}
\author{Shengping Gong}
\affil{Beihang University, Beijing, People's Republic of China}


\receiveddate{Manuscript received XXXXX 00, 0000; revised XXXXX 00, 0000; accepted XXXXX 00, 0000.\\
This work was supported by the National Natural Science Foundation of China (Grant No. 12372044), the National Natural Science Foundation of China (No. U23B6002), the National Natural Science Foundation of China (Grant No. 12302058), the National Natural Science Foundation of China (Grant No. 62227901), and the Postdoctoral Science Foundation of China (Grant No. 2024T170480).}

\corresp{{\itshape (Corresponding author: Shengping Gong)}.}

\authoraddress{Shuyue Fu is with the School of Astronautics and Shen Yuan Honors College, Beihang University, Beijing, 100191, People's Republic of China (e-mail: \href{fushuyue@buaa.edu.cn}{fushuyue@buaa.edu.cn}). Di Wu is also with the School of Astronautics, Beihang University, Beijing, 100191, People's Republic of China. He is also with the Key Laboratory of Spacecraft Design Optimization $\&$ Dynamic Simulation Technologies, Ministry of Education, Beijing, 100191, People's Republic of China (e-mail: \href{wudi2025@buaa.edu.cn}{wudi2025@buaa.edu.cn}). Xiaowen Liu is with the School of Astronautics, Beihang University, Beijing, 100191, People's Republic of China (e-mail: \href{Liuxiaowen2025@buaa.edu.cn}{Liuxiaowen2025@buaa.edu.cn}). Peng Shi is with the School of Astronautics, Beihang University, Beijing, 100191, People's Republic of China. He is also with the Key Laboratory of Spacecraft Design Optimization $\&$ Dynamic Simulation Technologies, Ministry of Education, Beijing, 100191, People's Republic of China (e-mail: \href{shipeng@buaa.edu.cn}{shipeng@buaa.edu.cn}). Shengping Gong is with the School of Astronautics, Beihang University, Beijing, 100191, People's Republic of China. He is also with the State Key Laboratory of High-Efficiency Reusable Aerospace Transportation Technology, Beijing, 102206, People's Republic of China (e-mail: \href{gongsp@buaa.edu.cn}{gongsp@buaa.edu.cn}).}

\markboth{FU ET AL.}{FOUR-BODY BALLISTIC LUNAR TRANSFERS USING ANALYTICAL ENERGY CONDITIONS}
\maketitle

\begin{abstract} 
The current lunar exploration, particularly the regular large-scale cargo transportation in the Earth-Moon system proposes requirements for amount of low-energy lunar transfers. The conventional grid-search method to construct low-energy lunar transfers suffers from extensive computational effort and large-scale searches. To further improve the method, this paper focuses on one type of low-energy lunar transfers termed ballistic lunar transfers, and derives prior knowledge to narrow the scale of searches. The Sun-Earth/Moon planar bicircular restricted four-body problem (PBCR4BP) is adopted as the dynamical model to construct lunar transfers. First, the analytical conditions for ballistic capture are derived and summarized in form of exact ranges of the Jacobi energy at the lunar insertion point. Both sufficient and necessary condition and necessary condition are developed. These conditions suggest an important role of the Sun-Earth/Moon PBCR4BP rather than the Earth-Moon planar restricted three-body problem in achieving lunar ballistic capture. Then, a grid-search method combined with the analytical energy conditions is proposed to construct ballistic lunar transfers. Simulations shows that a high ballistic capture ratio is achieved by the proposed method (99.87$\%$ for direct insertion and 98.72$\%$ for retrograde insertion). Examining the obtained ballistic lunar transfers, the effectiveness of the analytical energy conditions is verified. Samples of our obtained lunar transfers achieves lower or comparable impulses compared to solutions obtained in the previous works. Solutions belonging to new or less-reported transfer families are also presented, and the potential engineering applications of these trajectories are briefly discussed.  
\end{abstract}

\begin{IEEEkeywords}Planar restricted four-body problem, Lunar transfer, Lunar ballistic capture, Analytical energy conditions
\end{IEEEkeywords}

\section{INTRODUCTION}
I{\scshape nterest} in lunar transfers has increased with renewed lunar exploration missions. Transfers with less fuel consumption are particularly focused on due to the requirement of the regular large-scale cargo transportation in the Earth-Moon system to build a lunar infrastructure (e.g., \textit{Lunar Scientific Research Base} \cite{Pollpeter2023}). To construct lunar transfers satisfying mission requirements, scholars have proposed several types of transfers, such as low-energy impulsive transfers \cite{belbruno1993sun,Mingotti2012,Koon2001} and low-thrust transfers \cite{oshima2017global,lilow,du2024optimal}. Compared to continuous-thrust transfers, impulsive transfers are more mature, easier to implement, and widely adopted in Earth-Moon transfer missions (including low-energy transfer missions based on impulsive maneuvers, such as \textit{Hiten} \cite{belbruno1993sun} and \textit{Danuri} \cite{Song2023}). Therefore, we focus on the construction of low-energy transfers (based on impulsive maneuvers). For this type of transfer, bi-impulsive transfer is a classical problem and can serve as a preliminarily design of other multi-impulsive transfer \cite{topputo2013optimal,grossi2024optimal}. To construct low-energy bi-impulsive transfers, selecting appropriate dynamical models is important. Belbruno and Miller \cite{belbruno1993sun} pioneered low-energy transfers in the multi-body dynamics, verifying their solutions in the \textit{Lunar Observer} and \textit{Hiten} missions. Their work highlighted the role of multi-body gravity perturbation in constructing low-energy transfers. Sweetser \cite{bib51} theoretically demonstrated potential fuel savings when using the Earth-Moon planar circular restricted three-body problem (PCR3BP). Furthermore, Topputo \cite{topputo2013optimal} and Oshima et al. \cite{oshima2019low} pointed out that using the Sun-Earth/Moon planar bicircular restricted four-body problem (PBCR4BP) can further reduce impulses. Therefore, this Note focuses on the construction of low-energy lunar transfers in the Sun-Earth/Moon PBCR4BP.

Because closed-form solutions do not exist for the Sun-Earth/Moon PBCR4BP and other multi-body problems, the construction of low-energy transfers depends on numerical methods, mainly categorized into construction methods based on prior knowledge and grid-search methods \cite{dutt2018review,scheuerle2025energy}. Construction methods based on prior knowledge mainly focus on using weak stability boundary (WSB) theory \cite{belbruno1993sun,romagnoli2009earth,fu2026deep} and invariant manifolds of the libration periodic orbits \cite{Koon2001,belbruno2010weak,du2022low} (when considering transfers in the Sun-Earth/Moon PBCR4BP, the analogs of invariant manifolds termed Lagrangian coherent structures (LCSs) \cite{Onozaki2017,FU20254993,fu2025four} are also adopted). Meanwhile, other methods also utilized the dynamical structures of Moon collision trajectories \cite{oshima2017analysis} and those of double lunar gravity assist (LGA) trajectories \cite{qi2017optimal}. Notably, this type of method may be combined optimization method to obtain solutions with relatively low impulse \cite{Mingotti2012,qi2017optimal}. In this paper, we use ‘methods using prior knowledge’ to refer to those approaches where the initial guesses are generated from a perspective of prior knowledge or dynamical structures. This type of method uses natural dynamics to assist design. However, they typically yield fewer solutions than grid-search methods, which explore the solution space more broadly and allow more feasible solutions. Topputo \cite{topputo2013optimal} and Oshima et al. \cite{oshima2019low} employed grid-search method combined with continuation to obtain global maps of impulse and time of flight (TOF) for lunar transfers within 100 and 200 days, further exploring the transfer characterizations. In particular, Oshima et al. \cite{oshima2019low} pointed out that most of low-energy transfers end up with ballistic capture. Although grid-search method can explore the solution space of lunar transfer more broadly, its application often involves extensive computational effort and large-scale searches \cite{dutt2018review}. Furthermore, the Sun-Earth/Moon PBCR4BP introduces additional complexity due to increased parameter dimensionality caused by time-varying Sun-perturbed dynamics \cite{topputo2013optimal}. To address these challenges and construct low-energy transfers in a target way, it is motivated to utilize prior knowledge about multi-body dynamics to assist the construction of transfers by grid-search methods. Therefore, this paper explores this type of method. According to Refs. \cite{belbruno1993sun,Koon2001,topputo2013optimal,oshima2019low}, we find that ballistic capture plays an important role in low-energy transfers. Ballistic capture can reduce the Moon insertion impulse, therefore reducing the total impulse required. The aforementioned methods using grid-search methods mainly examine the ballistic capture after obtaining solutions. To use the effect of ballistic capture during the construction, we instead derive the analytical prior knowledge of ballistic capture and combine it with grid search. This method, however, still requires further exploration \cite{gagg2019method}. 

For lunar ballistic capture, several scholars developed the corresponding theory, both numerically and analytically. Belbruno \cite{belbruno2004capture} presented the approximation value of the Jacobi energy when the trajectory with respect to the Moon is a lunar parabolic trajectory. This approximation value was used to determine the energy-match relationship between the WSBs with respect to the Earth and the Moon. Furthermore, the way of lunar insertion of lunar transfers can have a further effect on the fuel consumption of low-energy transfers \cite{campagnola2010endgame,oshima2019low,FU20254993}. According to the way of lunar insertion, ballistic capture can be categorized into direct and retrograde capture. The numerical results obtained from Qi and Xu \cite{qi2014lunar} and Ano{\`e} et al. \cite{anoe2024ballistic} implied that there are specific ranges of the Jacobi energy for direct and retrograde capture around the Moon. Motivated by their results, we analytically derive energy conditions (necessary and sufficient conditions) for direct and retrograde capture at the lunar insertion point. Differing from the approximation value presented by Belbruno \cite{belbruno2004capture}, we obtain the closed-form expressions of the energy conditions, as constructing ballistic lunar transfers though grid-search methods (including trajectory correction) necessitates the exact ranges of the Jacobi energy. The obtained necessary and sufficient conditions for direct and retrograde capture are expressed in the form of the specific ranges of the Jacobi energy at the lunar insertion point. The lower boundaries of the Jacobi energy for direct and retrograde capture are equivalent to the expressions of the Jacobi energy developed by Fantino et al. \cite{fantino2010note} when the eccentricity with respect to the Moon is set to 1. However, the monotonicity between the Jacobi energy and eccentricity (equivalent to the Keplerian energy with respect to the Moon) has not been thoroughly analyzed, and the energy conditions for ballistic capture have not been comprehensively investigated. We perform this analysis in terms of the Jacobi energy and Keplerian energy with respect to the Moon, deriving the necessary and sufficient conditions for ballistic capture. These conditions also suggest the important role of the solar gravity perturbation introduced to the Earth-Moon PCR3BP (i.e., the Sun-Earth/Moon PBCR4BP) in achieving lunar ballistic capture, and this can be considered as a theoretical contribution made in this paper. Then, to use these energy conditions for construction of ballistic lunar transfers, we weaken the necessary and sufficient conditions to the necessary conditions, and propose a grid-search method combined with these conditions. Because the necessary conditions are expressed in terms of the states at the insertion point, backward time propagation is employed to construct lunar transfers. In the obtained solutions, the ratio of ballistic capture is $1589/1591$ (99.87$\%$) for direct insertion and $386/391$ (98.72$\%$) for retrograde insertion. Meanwhile, the values of the Jacobi energy at the insertion point are consistent with the derived analytical energy conditions. Our obtained solutions with the minimum impulses and ballistic capture (direct and retrograde capture) are lower than or comparable to those obtained from the conventional methods \cite{belbruno1993sun,Mingotti2012,Onozaki2017,oshima2017analysis,qi2017optimal,topputo2013optimal,oshima2019low} (with the same constraints). Meanwhile, trajectories belonging to new or less-reported transfer families are found in the obtained solutions compared to the previous works \cite{topputo2013optimal,oshima2019low,oshima2017analysis,qi2017optimal,campana2024clustering,campana2025ephemeris}. These trajectories move in tadpole-like orbits \cite{oshima2015jumping} in the L4 region before achieving ballistic capture. These type of trajectories may be applicable to a single-launch dual-mission framework: one mission for lunar exploration, and the other one mission for the Earth-Moon L4 exploration. Based on the aforementioned discussion, main contributions of this paper can be summarized as follows:
\begin{enumerate}
  \item We derive and summarize analytical energy conditions for ballistic capture, providing a theoretical supplementation to previous works \cite{belbruno2004capture,fantino2010note,oshima2019low,anoe2024ballistic}. The derived energy conditions suggest the important role of the solar gravity perturbation in achieving lunar ballistic capture.
  \item We propose a grid-search method combined with the energy conditions to construct ballistic lunar transfers in a target way. Simulations show that our solutions achieve a high ballistic capture ratio. Compared to solutions obtained from conventional methods only using prior knowledge \cite{belbruno1993sun,Mingotti2012,Onozaki2017,oshima2017analysis,qi2017optimal}, samples of our obtained solutions achieve lower or comparable impulses.
  \item Trajectories belonging to new or less-reported transfer families compared to the previous works \cite{topputo2013optimal,oshima2019low,oshima2017analysis,qi2017optimal,campana2024clustering,campana2025ephemeris} are found, providing a potential engineering application for a single-launch dual-mission framework.
\end{enumerate}

The rest of this paper is organized as follows. Section \ref{sec2} presents the scenario description of lunar transfers in the Sun-Earth/Moon PBCR4BP. Section \ref{sec3} derives the analytical energy conditions for ballistic capture. Section \ref{sec4} proposes the method to construct the ballistic lunar transfers and analyze the results. Finally, conclusions are drawn in Section \ref{sec5}.

\section{LUNAR TRANSFER IN THE SUN-EARTH/MOON PBCR4BP}\label{sec2}
This section describes the transfer scenario in the Sun-Earth/Moon PBCR4BP, including the dynamical model and lunar transfers. 
\subsection{Sun-Earth/Moon PBCR4BP}\label{subsec2.3}
The Sun-Earth/Moon PBCR4BP is adopted to construct lunar transfers \cite{topputo2013optimal,fu2025four}. When describing the dynamical equations, the dimensionless units are selected as follows: the length unit (LU) is set as the Earth-Moon distance, the mass unit (MU) is set as the combined mass of the Earth and Moon, and the time unit (TU) is set as ${\text{TU}}={T_\text{EM}}/{2\pi}$, where ${T_{{\text{EM}}}}$ is the orbital period of the Earth and Moon about their barycenter. The Earth-Moon rotating frame \cite{topputo2013optimal} is adopted, and the dynamical equations are expressed as:
\begin{equation}
\left[ {\begin{array}{*{20}{c}}
{\begin{array}{*{20}{c}}
{\dot x}\\
{\dot y}
\end{array}}\\
{\begin{array}{*{20}{c}}
{\dot u}\\
{\dot v}
\end{array}}
\end{array}} \right] = \left[ {\begin{array}{*{20}{c}}
{\begin{array}{*{20}{c}}
u\\
v
\end{array}}\\
{\begin{array}{*{20}{c}}
{2v + \frac{{\partial {\Omega _4}}}{{\partial x}}}\\
{ - 2u + \frac{{\partial {\Omega _4}}}{{\partial y}}}
\end{array}}
\end{array}} \right]\label{eq_PBCR4BP}
\end{equation}
\begin{align}
{\Omega _4} &= \frac{1}{2}\left[ {{x^2} + {y^2} + \mu \left( {1 - \mu } \right)} \right] + \frac{{1 - \mu }}{{{r_1}}} + \frac{\mu }{{{r_2}}} + \frac{{{\mu_{\text{S}}}}}{{{r_3}}} \\ \notag &- \frac{{{\mu_{\text{S}}}}}{{{\rho ^2}}}\left( {x\cos {\theta _{\text{S}}} + y\sin {\theta _{\text{S}}}} \right)\label{eq_omega4}
\end{align}
where $\bm{X} = \left[ x, \text{ }y, \text{ }u,\text{ } v\right]^{\text{T}}$ denotes the orbital state, $\Omega _4$ denotes the effective potential of the PBCR4BP, $\mu$ denotes the mass parameter, and the parameter $\mu_{\text{S}}$ denotes the dimensionless mass of the Sun. The distances between the spacecraft and the Earth ($r_1$), Moon ($r_2$), and Sun ($r_3$) are expressed as:
\begin{equation}
{r_1} = \sqrt {{{\left( {x + \mu } \right)}^2} + {y^2}}   \label{eq_r1}
\end{equation}
\begin{equation}
{r_2} = \sqrt {{{\left( {x + \mu - 1} \right)}^2} + {y^2}} \label{eq_r2}
\end{equation}
\begin{equation}
{r_3} = \sqrt {{{\left( {x - \rho \cos {\theta _{\text{S}}}} \right)}^2} + {{\left( {y - \rho \sin {\theta _{\text{S}}}} \right)}^2}}  \label{eq_r3}
\end{equation}
where $\rho$ denotes the distance between the Earth-Moon barycenter and the Sun, and the solar phase angle ${\theta _{\text{S}}}$ is calculated by ${\theta _{\text{S}}} = {\theta _{{\text{S0}}}} + {\omega _{\text{S}}}T$, where ${\theta _{\text{S0}}}$ denotes the initial solar phase angle expressed as ${\theta _{\text{S0}}}={\omega _{\text{S}}}t_0$ and $T$ denotes the propagation time \cite{FU20254993}. For numerical propagation of trajectories in the Sun-Earth/Moon PBCR4BP, we use the MATLAB®'s ode113 command with absolute and relative tolerances set to $1 \times 10^{-13}$ \cite{oshima2021capture}. Specific values of the parameters used in simulations can be found in Refs. \cite{FU20254993,fu2025four}.

In the Sun-Earth/Moon PBCR4BP, the Jacobi energy is time-varying because of Sun-perturbed dynamics. According to Ref. \cite{FU20254993}, the instantaneous Jacobi energy is defined as:
\begin{equation}
C = -\left( {{u^2} + {v^2}} \right) +  \left( {{x^2} + {y^2}} \right) + \frac{{2(1 - \mu) }}{{{r_1}}} + \frac{2\mu }{{{r_2}}} + \mu \left( {1 - \mu } \right) \label{eq4}
\end{equation}

Subsequently, the concepts of the bi-impulsive lunar transfers, ballistic capture, and ballistic lunar transfers are introduced.
\subsection{Lunar Transfers and Ballistic Lunar Transfers}\label{subsec2.2}
In this paper, we focus on bi-impulsive lunar transfers, particularly those with ballistic capture (i.e., ballistic lunar transfers). In such transfers, the spacecraft departs a circular Earth parking orbit with an Earth injection impulse ($\Delta {v_i}$) and inserts into a circular lunar insertion orbit after performing a Moon insertion impulse ($\Delta {v_f}$). The impulses should be tangential to the orbital velocity to maximize energy variations \cite{bib49}. With dimensionless units, the constraints of lunar transfers can be expressed as \cite{topputo2013optimal,oshima2019low}:
\begin{equation}
{\bm{\psi }_i} = \left[ {\begin{array}{*{20}{c}}
  {{{\left( {{x_i} + \mu } \right)}^2} + {y_i}^2 - {{\left( {{R_{\text{E}}} + {h_i}} \right)}^2}} \\ 
  {\left( {{x_i} + \mu } \right)\left( {{u_i} - {y_i}} \right) + {y_i}\left( {{v_i} + {x_i} + \mu } \right)} 
\end{array}} \right] = \mathbf{0} \label{eq5}
\end{equation}
\begin{equation}
{\bm{\psi }_f} = \left[ {\begin{array}{*{20}{c}}
  {{{\left( {{x_f} + \mu  - 1} \right)}^2} + {y_f}^2 - {{\left( {{R_{\text{M}}} + {h_f}} \right)}^2}} \\ 
  {\left( {{x_f} + \mu  - 1} \right)\left( {{u_f} - {y_f}} \right) + {y_f}\left( {{v_f} + {x_f} + \mu  - 1} \right)} 
\end{array}} \right] = \mathbf{0} \label{eq6}
\end{equation}
where $h_i$ and $h_f$ denote altitudes of the Earth parking orbit and lunar insertion orbit. The subscript ‘\textit{i}’ and ‘\textit{f}’ denote quantities associated with the departure and insertion points. These constraints ensure that the transfer trajectories depart from the Earth parking orbit using a tangential impulse in the Earth-centered inertial frame and insert into the lunar insertion orbit using a tangential impulse in the Moon-centered inertial frame. With these constraints, we present the definition of bi-impulsive lunar transfers:
\begin{definition}[Bi-impulsive Lunar Transfer]\label{def2.0}
Bi-impulsive lunar transfer is the transfer trajectory that satisfies Eqs. \eqref{eq5}-\eqref{eq6}, where $\bm{X}_f=\phi _{{t_i}}^{{t_f}}\left( {{\bm{X}_i}} \right)$ and $\phi _{{t_i}}^{{t_f}}:\mathbb{R} \times \mathbb{R} \times {\text{D}} \to {\text{D;   }}\left( {{t_i},{\text{ }}{t_f},{\text{ }}{\bm{X}_i}} \right) \to \phi _{{t_i}}^{{t_f}}\left( {{\bm{X}_i}} \right)$ is the flow map of the Sun-Earth/Moon PBCR4BP. The parameters $t_i$ and $t_f$ denote the departure and insertion epochs.
\end{definition}

In this paper, we set $h_i$ and $h_f$ to 167 km and 100 km \cite{topputo2013optimal,oshima2019low}, respectively. In the following text, we denote ${{ {{R_{\text{E}}} + {h_i}}}}$ and ${{ {{R_{\text{M}}} + {h_f}} }}$ as $r_i$ and $r_f$. Then, the impulses of the transfer trajectory can be calculated by:
\begin{equation}
\Delta {v_i} = \sqrt {{{\left( {{u_i} - {y_i}} \right)}^2} + {{\left( {{v_i} + {x_i} + \mu } \right)}^2}}  - \sqrt {\frac{{1 - \mu }}{r_i}}  \label{eq7}
\end{equation}
\begin{equation}
\Delta {v_f} = \sqrt {{{\left( {{u_f} - {y_f}} \right)}^2} + {{\left( {{v_f} + {x_f} + \mu  - 1} \right)}^2}}  - \sqrt {\frac{\mu }{r_f}}   \label{eq8}
\end{equation}
\begin{equation}
\Delta {v} = \Delta {v_i}+\Delta {v_f}  \label{eq_dv}
\end{equation}
where $\Delta {v}$ denotes the total impulse. In bi-impulsive lunar transfers, ballistic capture (i.e., negative Keplerian energy with respect to the Moon) at the insertion point plays an important role in reducing $\Delta{v}_f$, which consequently leads to the reduction of $\Delta {v}$ \cite{belbruno1993sun}. Moreover, way of ballistic capture (i.e., direct and retrograde capture) can further affect $\Delta{v}_f$ \cite{campagnola2010endgame,oshima2019low,FU20254993}. Therefore, both direct and retrograde capture are investigated. To define direct and retrograde capture, two parameters at the insertion point are introduced, namely, the Keplerian energy with respect to the Moon ($E_f$) and the angular momentum with respect to the Moon ($M_f$). The Keplerian energy with respect to the Moon at the insertion point is expressed as:
\begin{equation}
{E_f} = \frac{1}{2}\left[ {{{\left( {{u_f} - {y_f}} \right)}^2} + {{\left( {{v_f} + {x_f} + \mu  - 1} \right)}^2}} \right] - \frac{\mu }{r_f} \label{eq9}
\end{equation}
With the expression of the Keplerian energy with respect to the Moon at the insertion point, we present the definition of the ballistic capture at the insertion point \cite{topputo2013optimal}:
\begin{definition}[Ballistic Capture]\label{def2.1}
Ballistic capture at the insertion point is the state that satisfies $E_f \leq 0$.
\end{definition}
Notably, there have also been several definitions of ballistic capture in previous works \cite{hyeraci2013role,luo2020role,anoe2024ballistic}. Moreover, there have also several capture strategies, including direct (impulsive) capture, aerocapture, and low-thrust capture \cite{zhang2025noncoplanar,yao2022nonsingular,falcone2022autonomous}. Here, we adopt the definition in Ref. \cite{topputo2013optimal} and the direct capture strategy, which are suitable for the lunar transfer scenario. According to Definition \ref{def2.1}, $E_f$ is the key parameter to identify transfer trajectories with ballistic capture. Then, the angular momentum with respect to the Moon at the insertion point is expressed as:
\begin{equation}
{M_f} = \left( {{x_f} + \mu  - 1} \right)\left( {{v_f} + {x_f} + \mu  - 1} \right) - {y_f}\left( {{u_f} - {y_f}} \right) \label{eq10}
\end{equation}
According to the signs of $M_f$, direct and retrograde capture can be defined as:
\begin{definition}[Direct Capture]\label{def2.2}
Direct capture at the insertion point is ballistic capture that satisfies $M_f > 0$.
\end{definition}
\begin{definition}[Retrograde Capture]\label{def2.3}
Retrograde capture at the insertion point is ballistic capture that satisfies $M_f < 0$.
\end{definition}
Then, with the definitions of bi-impulsive lunar transfers and ballistic capture, we present the definition of ballistic lunar transfers:
\begin{definition}[Ballistic Lunar Transfer]\label{def2.4}
Ballistic lunar transfer is the transfer trajectory that satisfies Definition \ref{def2.0} and ends up with ballistic capture defined in Definition \ref{def2.1}.
\end{definition}

Subsequently, to construct ballistic lunar transfers, we derive the analytical energy conditions for ballistic capture.

\section{ANALYTICAL ENERGY CONDITIONS FOR BALLISTIC CAPTURE}\label{sec3}
In this section, we derive the analytical energy conditions for ballistic capture at the insertion point. Firstly, we address the critical cases to achieve direct and retrograde capture, i.e., the cases where $E_f=0$. Then, we summarize the analytical energy conditions for ballistic capture.
\subsection{Critical Case: $E_f=0$}\label{subsec3.1}
We start with the critical case, i.e., $E_f=0$. According to Ref. \cite{anoe2024ballistic}, $E_f$ and $M_f$ have a relationship shown as follows:
\begin{align}\label{eq11}
C_f &=  - 2{E_f} + 2{M_f} + 2\left( {1 - \mu } \right){x_f} \\ 
\notag &+ \left( {1 - \mu } \right)\left( {2\mu  - 1} \right) + \frac{{2\left( {1 - \mu } \right)}}{{{r_{1f}}}} 
\end{align}
where $C_f$ denotes the Jacobi energy at the insertion point. When $E_f=0$, Eq. \eqref{eq11} can be simplified into:
\begin{align}\label{eq11_new}
C_f &=  2{M_f} + 2\left( {1 - \mu } \right){x_f} + \left( {1 - \mu } \right)\left( {2\mu  - 1} \right) \\ \notag & + \frac{{2\left( {1 - \mu } \right)}}{{{r_{1f}}}}
\end{align}
Subsequently, based on this relationship, we derive the analytical energy conditions for ballistic capture.
\subsubsection{Direct Insertion}\label{subsec3.1.1}
\begin{figure*}
\centerline{\includegraphics[width=23pc]{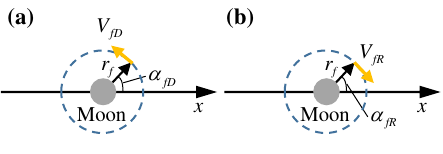}}
\caption{The definition of the phase angle. (a) Direct insertion; (b) Retrograde insertion.}
\label{fig_alpha}
\end{figure*}

To analyze the critical case to achieve direct capture, we firstly parameterize the states at the insertion point under the specific value of Jacobi energy. According to Eq. \eqref{eq6}, the states with $M_f>0$ can be parameterized as:
\begin{equation}\label{eq13}
\left\{ \begin{gathered}
  {x_{fD}} = {r_f}\cos {\alpha _{fD}} + 1 - \mu  \hfill \\
  {y_{fD}} = {r_f}\sin {\alpha _{fD}} \hfill \\
  {u_{fD}} =  - {V_{fD}}\sin {\alpha _{fD}} \hfill \\
  {v_{fD}} = {V_{fD}}\cos {\alpha _{fD}} \hfill \\ 
\end{gathered}  \right.
\end{equation}
\begin{equation}\label{eq13_new}
{V_{fD}} = \sqrt {\begin{gathered}
   - {C_{fD}} + \left( {{x_{fD}}^2 + {y_{fD}}^2} \right) + \frac{{2\left( {1 - \mu } \right)}}{{{r_{1fD}}}}  \\
   + \frac{{2\mu }}{{{r_f}}} + \mu \left( {1 - \mu } \right)  \\ 
\end{gathered} }
\end{equation}
where $\alpha_f$ denotes the phase angle (see Fig. \ref{fig_alpha} (a)), and the subscript ‘\textit{D}’ denotes quantities associated with direct insertion (including direct capture). 

For direct insertion, the angular momentum with respect to the Moon at the insertion point can be obtained by substituting Eq. \eqref{eq13} into Eq. \eqref{eq10}:
\begin{align}\label{eq15}
{M_{fD}} &= \left( {{x_{fD}} + \mu  - 1} \right)\left( {{v_{fD}} + {x_{fD}} + \mu  - 1} \right) \\ \notag& - {y_{fD}}\left( {{u_{fD}} - {y_{fD}}} \right)\\ \notag& ={r_f}^2+{r_f}{V_{fD}} 
\end{align}
where
\begin{equation}
W\left({\alpha _{f}}\right) =  \left( {{x_{f}}^2 + {y_{f}}^2} \right) + \frac{{2\left( {1 - \mu } \right)}}{{{r_{1f}}}} + \frac{{2\mu }}{{{r_{f}}}} + \mu \left( {1 - \mu } \right) \label{eq16}
\end{equation}
According to the relationship shown in Eq. \eqref{eq11_new}, we consider the critical case, i.e., $E_{fD}=0$, and reconstruct Eq. \eqref{eq11_new} as:
\begin{align}\label{eq17}
{C_{fD}}^* &= 2{r_f}^2 + 2{r_f}\sqrt { - {C_{fD}}^* + W\left( {{\alpha _{fD}}} \right)} \\ \notag & + 2\left( {1 - \mu } \right)\left( {{r_f}\cos {\alpha _{fD}} - \mu  + 1} \right)\\ \notag & + \left( {1 - \mu } \right)\left( {2\mu  - 1} \right) + \frac{{2\left( {1 - \mu } \right)}}{{{r_{1fD}}}} 
\end{align}
where ${C_{fD}}^*$ denotes the critical Jacobi energy satisfying $E_{fD}=0$ for direct insertion. Rewriting Eq. \eqref{eq17} as:
\begin{equation}
{C_{fD}}^* = G\left({\alpha _{fD}}\right) + 2{r_f}\sqrt { - {C_{fD}}^* + W\left( {{\alpha _{fD}}} \right)} \label{eq18}
\end{equation}
where
\begin{align}\label{eq19}
G\left({\alpha _{f}}\right) &=  2{r_{f}}^2+ 2\left( {1 - \mu } \right)\left( {{r_{f}}\cos {\alpha _{f}} - \mu  + 1} \right) \\ \notag & + \left( {1 - \mu } \right)\left( {2\mu  - 1} \right) + \frac{{2\left( {1 - \mu } \right)}}{{{r_{1f}}}}
\end{align}

Therefore, the equations that ${C_{fD}}^*$ should satisfy can be expressed as:

\begin{equation}
\begin{gathered}
  {{C_{fD}}^*}^2 - 2\left( {G\left( {{\alpha _{fD}}} \right) - 2{r_f}^2} \right){C_{fD}}^* \hfill \\
   + \left( {{G^2}\left( {{\alpha _{fD}}} \right) - 4{r_f}^2W\left( {{\alpha _{fD}}} \right)} \right) = 0
   \hfill \\
\end{gathered}
\label{eq20}
\end{equation}

\begin{equation}
\begin{gathered}
  2{r_f}\sqrt { - {C_{fD}}^* + W\left( {{\alpha _{fD}}} \right)}  = {C_{fD}}^* - G\left( {{\alpha _{fD}}} \right) \geq 0 \hfill \\
   \Rightarrow {C_{fD}}^* \geq G\left( {{\alpha _{fD}}} \right) \hfill \\ 
\end{gathered} 
\label{eq20_1}
\end{equation}
Equation \eqref{eq20} is a quadratic equation. The discriminant of this equation can be calculated by:
\begin{align} \label{eq21}
  \Delta & = 4{\left( {G\left( {{\alpha _{fD}}} \right) - 2{r_f}^2} \right)^2} \\ \notag &  - 4\left( {{G^2}\left( {{\alpha _{fD}}} \right) - 4{r_f}^2W\left( {{\alpha _{fD}}} \right)} \right)\\ \notag &  = 16{r_f}^2\left( {{r_f}^2 - G\left( {{\alpha _{fD}}} \right) + W\left( {{\alpha _{fD}}} \right)} \right) \\ \notag &  = 32\mu {r_f} > 0 
\end{align}
Because $\Delta>0$, there exist two solutions to the equation. Due to ${C_{fD}}^* \geq G\left( {{\alpha _{fD}}} \right)$, the value of ${C_{fD}}^*$ is solved by:
\begin{align}\label{eq22}
{C_{fD}}^* & = \frac{2\left(G\left( {{\alpha _{fD}}} \right)-2{r_f}^2\right)+\sqrt{\Delta}}{2} \\ \notag & = G\left( {{\alpha _{fD}}} \right) - 2{r_f}^2 + 2\sqrt {2\mu {r_f}} 
\end{align}
According to Eq. \eqref{eq22}, the value of ${C_{fD}}^*$ depends on the value of ${{\alpha _{fD}}}$. Then, the minimum value of ${C_{fD}}^*$ with respect to $\alpha_{fD}$ is analyzed. The derivative of ${C_{fD}}^*$ with respect to $\alpha_{fD}$ is expressed as:
\begin{align}\label{eq25}
\frac{{{\text{d}}{C_{fD}}^*\left( {{\alpha _{fD}}} \right)}}{{{\text{d}}{\alpha _{fD}}}} & =  - 2\left( {1 - \mu } \right){r_f}\sin {\alpha _{fD}} \\ \notag & + \frac{{2\left( {1 - \mu } \right){r_f}\sin {\alpha _{fD}}}}{{{r_{1fD}}^3}}
\end{align}
The derivative of ${C_{fD}}^*$ with respect to $\alpha_{fD}$ has five zero points, expressed as follows:
\begin{equation}
\begin{gathered}
  {\alpha _{fD}}_1 = 0,{\text{  }}{\alpha _{fD}}_2 = \arccos \left( { - \frac{{{r_f}}}{2}} \right),{\text{  }}{\alpha _{fD}}_3 = \pi , \hfill \\
  {\alpha _{fD}}_4 = 2\pi  - \arccos \left( { - \frac{{{r_f}}}{2}} \right),{\text{  }}{\alpha _{fD}}_5 = 2\pi  \hfill \\ 
\end{gathered} \label{eq26}
\end{equation}
The value of ${C_{fD}}^*$ has a global minimum at ${\alpha _{fD}}_2$ and ${\alpha _{fD}}_4$. The minimum value of ${C_{fD}}^*$ is expressed as:
\begin{equation}
{{C_{fD}}^*{\left( {{\alpha _{fD}}} \right)_{\min }} = 3\left( {1 - \mu } \right) - \left( {1 - \mu } \right){r_f}^2 + 2\sqrt {2\mu {r_f}}} \label{eq28}
\end{equation}
Subsequently, we investigate the case of retrograde insertion.

\subsubsection{Retrograde Insertion}\label{subsec3.1.2}
The states with $M_f<0$ can be parameterized as:
\begin{equation}\label{eq14}
\left\{ \begin{gathered}
  {x_{fR}} = {r_f}\cos {\alpha _{fR}} + 1 - \mu  \hfill \\
  {y_{fR}} = {r_f}\sin {\alpha _{fR}} \hfill \\
  {u_{fR}} = {V_{fR}}\sin {\alpha _{fR}} \hfill \\
  {v_{fR}} =  - {V_{fR}}\cos {\alpha _{fR}} \hfill \\ 
\end{gathered}  \right.
\end{equation}
\begin{equation}\label{eq14_new}
{V_{fR}} = \sqrt {\begin{gathered}
   - {C_{fR}} + \left( {{x_{fR}}^2 + {y_{fR}}^2} \right) + \frac{{2\left( {1 - \mu } \right)}}{{{r_{1fR}}}}  \\
   + \frac{{2\mu }}{{{r_f}}} + \mu \left( {1 - \mu } \right)  \\ 
\end{gathered}}
\end{equation}
where the subscript ‘\textit{R}’ denotes quantities associated with retrograde insertion (including retrograde capture), and $\alpha_{fR}$ is shown in Fig. \ref{fig_alpha} (b). For retrograde insertion, the angular momentum with respect to the Moon at the insertion point can be obtained by substituting Eq. \eqref{eq14} into Eq. \eqref{eq10}:
\begin{align}\label{eq29}
{M_{fR}} & = \left( {{x_{fR}} + \mu  - 1} \right)\left( {{v_{fR}} + {x_{fR}} + \mu  - 1} \right) \\ \notag & - {y_{fR}}\left( {{u_{fR}} - {y_{fR}}} \right) \\ \notag & ={r_f}^2-{r_f}{V_{fR}} 
\end{align}
Similarly, considering the critical case $E_{fR}=0$, the equations requiring ${C_{fR}}^*$ to satisfy are presented as follows:
\begin{equation}
{C_{fR}}^* = G\left({\alpha _{fR}}\right) - 2{r_f}\sqrt { - {C_{fR}}^* + W\left( {{\alpha _{fR}}} \right)} \label{eq30}
\end{equation}

\begin{equation}
\begin{gathered}
  {{{C_{fR}}^*}^2} - 2\left( {G\left( {{\alpha _{fR}}} \right) - 2{r_f}^2} \right){C_{fR}}^* \hfill \\
   + \left( {{G^2}\left( {{\alpha _{fR}}} \right) - 4{r_f}^2W\left( {{\alpha _{fR}}} \right)} \right) = 0
   \hfill \\
\end{gathered}
\label{eq31}
\end{equation}

\begin{equation}
\begin{gathered}
  2{r_f}\sqrt { - {C_{fR}}^* + W\left( {{\alpha _{fR}}} \right)}  = -{C_{fR}}^* + G\left( {{\alpha _{fR}}}  \right) \geq 0 \hfill \\
   \Rightarrow {C_{fR}}^* \leq G\left( {{\alpha _{fR}}} \right) \hfill \\ 
\end{gathered} 
\label{eq31_1}
\end{equation}
Equation \eqref{eq31} has the same form as Eq. \eqref{eq20} but with different inequalities. Then, ${C_{fR}}^*$ is solved by:
\begin{align}\label{eq32}
{C_{fR}}^* & = \frac{2\left(G\left( {{\alpha _{fR}}} \right)-2{r_f}^2\right)-\sqrt{\Delta}}{2} \\ \notag & = G\left( {{\alpha _{fR}}} \right) - 2{r_f}^2 - 2\sqrt {2\mu {r_f}} 
\end{align}
Similar to the analysis of ${C_{fD}}^*$, the minimum value of ${C_{fR}}^*$ is presented as follows:
\begin{equation}
{C_{fR}}^*{\left( {{\alpha _{fR}}} \right)_{\min }} = 3\left( {1 - \mu } \right) - \left( {1 - \mu } \right){r_f}^2 - 2\sqrt {2\mu {r_f}} \label{eq35}
\end{equation}
\subsection{Analytical Conditions for Ballistic Capture}\label{subsec3.3}
\begin{figure*}
\centerline{\includegraphics[width=23pc]{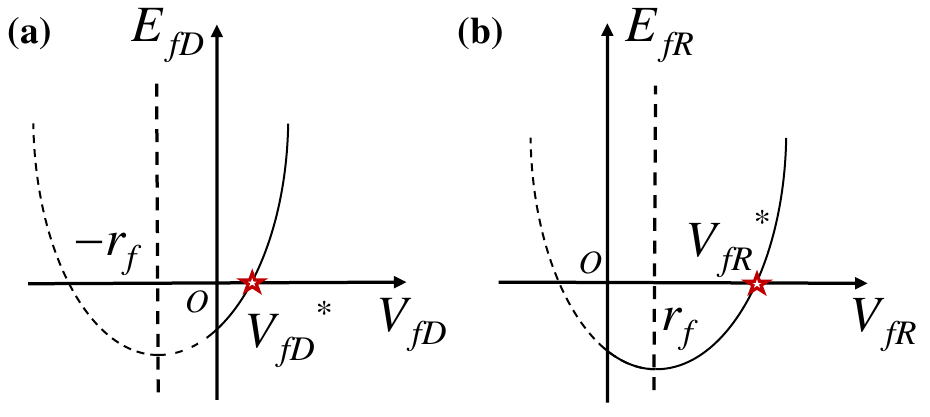}}
\caption{The schematic of $E_{f}$ as a quadratic function of ${V_{f}}$. (a) Direct insertion; (b) Retrograde insertion.}\label{fig_C_New}
\end{figure*}

When the values of $C_f$ for the critical cases are derived, we investigate the monotonicity between $C_f$ and $E_f$ to obtain the analytical energy conditions for ballistic capture. Following Section III-\ref{subsec3.1}, we illustrate that for each $\alpha_{f}$, when $C_{f}$ satisfies $C_{f}<{C_{f}}^*\left( {{\alpha _{f}}} \right)$, the Keplerian energy with respect to the Moon at the insertion point satisfies $E_{f}>0$. The expression of $E_{f}$ can be expressed as:
\begin{align} \label{eq23}
   {E_{f}} &= \frac{1}{2}\left[ {{{\left( {{u_{f}} - {y_{f}}} \right)}^2} + {{\left( {{v_{f}} + {x_{f}} + \mu  - 1} \right)}^2}} \right] - \frac{\mu }{{{r_f}}} \\ \notag & = \frac{1}{2}{\left( {\sqrt { - {C_{f}} + W\left( {{\alpha _{f}}} \right)}  \pm {r_f}} \right)^2} - \frac{\mu }{{{r_f}}}
\end{align}
where $+$ for direct insertion and $-$ for retrograde insertion. In this expression, we can observe that $E_{f}$ is a quadratic function of ${V_{f}}={\sqrt { - {C_{f}} + W\left( {{\alpha _{f}}} \right)}}$. For each $\alpha_{f}$, when $C_{f}={C_{f}}^*$, $V_{f}{={V_{f}}^*=\sqrt { - {C_{f}}^* + W\left( {{\alpha _{f}}} \right)}}$, the value of $E_{f}$ satisfies:
\begin{equation}
{{E_{f}}\left( {{C_{f}}^*\left( {{\alpha _{f}}} \right),{\text{ }}{\alpha _{f}}} \right) = 0} \label{eq24}
\end{equation}
The schematic of $E_{f}$ as a quadratic function of ${V_{f}}$ is presented in Fig. \ref{fig_C_New} (note that ${V_{f}}\geq 0$).

Meanwhile, $C_f$ should satisfy $C_{f} \leq W{\left( {{\alpha _{f}}} \right)}$ to ensure that $V_f$ is a real number. The aforementioned derivation is bidirectional, i.e., when $C_{f}$ satisfies ${C_{f}}^*{\left( {{\alpha _{f}}} \right)}\leq C_{f} \leq W{\left( {{\alpha _{f}}} \right)}$, $E_f$ satisfies $E_f \leq 0$; when $E_f \leq 0$, $C_f$ satisfies ${C_{f}}^*{\left( {{\alpha _{f}}} \right)}\leq C_{f} \leq W{\left( {{\alpha _{f}}} \right)}$. Therefore, we summarize the necessary and sufficient condition for ballistic capture at the insertion point:

\begin{theorem}[Sufficient and Necessary Condition for Ballistic Capture]\label{th3.1}
Ballistic capture at the insertion point $\left(x_{f},\text{ }y_{f}\right)$ takes place if the Jacobi energy at the insertion point $C_{f}$ satisfies ${C_{f}}^*{\left( {{\alpha _{f}}} \right)}\leq C_{f} \leq W{\left( {{\alpha _{f}}} \right)}$ (the converse is also true), where:
\begin{enumerate}
  \item $\alpha _{f}=\text{atan2}\left(y_{f},\text{ }x_{f}+\mu-1\right)$;
  \item ${C_{f}}^*{\left( {{\alpha _{f}}} \right)} = \left( {1 - \mu } \right) + 2\left( {1 - \mu } \right){r_f}{\cos}{\alpha _{f}}+\frac{2\left( {1 - \mu } \right)}{r_{1f}} \pm 2\sqrt {2\mu {r_f}}$ ($+$ for direct capture and $-$ for retrograde capture);
  \item $W\left({\alpha _{f}}\right) =  \left( {{x_{f}}^2 + {y_{f}}^2} \right) + \frac{{2\left( {1 - \mu } \right)}}{{{r_{1f}}}} + \frac{{2\mu }}{{{r_{f}}}} + \mu \left( {1 - \mu } \right)$.
\end{enumerate}
\end{theorem}
Here we present the proof of this theorem (taking the direct capture for example):
\begin{proof}[Proof of Theorem \ref{th3.1}]\label{proof3.1} (Proof of Sufficiency) Sufficiency of Theorem \ref{th3.1} can be proved by Eqs. \eqref{eq15}-\eqref{eq24}. When $C_{fD}\geq{C_{fD}}^*$, i.e., $V_{fD}\leq{V_{fD}}^*$, $E_{fD}$ satisfies $E_{fD}\leq0$, as shown in Fig. \ref{fig_C_New} (a). Meanwhile $V_{fD}\geq 0$, thus $C_{fD}\leq W{\left( {{\alpha _{fD}}} \right)}$. Then, we have:
\begin{equation}
{C_{fD}}^*\left( {{\alpha _{fD}}} \right) \leq {C_{fD}} \leq W\left( {{\alpha _{fD}}} \right) \Rightarrow {E_{fD}} \leq 0
\end{equation}
(Proof of Necessity) As shown in Fig. \ref{fig_C_New} (a), when ${E_{fD}} \leq 0$, we have $0\leq V_{fD} \leq {V_{fD}}^*$. Because $V_{fD}$ and $C_{fD}$ have a relationship $V_{fD}=\sqrt{-C_{fD}+W\left( {{\alpha _{fD}}} \right)}$, $C_{fD}$ should satisfy ${C_{fD}}^*{\left( {{\alpha _{fD}}} \right)}\leq C_{fD} \leq W{\left( {{\alpha _{fD}}} \right)}$. Therefore, we have:
\begin{equation}
{E_{fD}} \leq 0 \Rightarrow {C_{fD}}^*\left( {{\alpha _{fD}}} \right) \leq {C_{fD}} 
\end{equation}

\end{proof}

To apply the analytical energy conditions to constructing ballistic lunar transfers in the Sun–Earth/Moon PBCR4BP and to facilitate a grid search for each $\alpha_f$, we weaken the aforementioned necessary and sufficient condition for direct and retrograde capture to the following necessary condition:

\begin{theorem}[Necessary Condition for Ballistic Capture]\label{th3.3}
The Jacobi energy at the insertion point $C_{f}$ satisfies ${C_{f}}^*{\left( {{\alpha _{f}}} \right)_{\min}}\leq C_{f} \leq W{\left( {{\alpha _{f}}} \right)_{\max}}$ if ballistic capture at the insertion point $\left(x_{f},\text{ }y_{f}\right)$ takes place, where ${C_{f}}^*{\left( {{\alpha _{f}}} \right)_{\min }} = 3\left( {1 - \mu } \right) - \left( {1 - \mu } \right){r_f}^2 \pm 2\sqrt {2\mu {r_f}}$ ($+$ for direct capture and $-$ for retrograde capture).
\end{theorem}

The proof of the necessary condition is apparent. For lunar transfers considered in this paper ($h_f=100\text{ km}$), the exact values of ${C_{fD}}^*\left(\alpha_{fD}\right)_{\min}$ and ${C_{fR}}^*\left(\alpha_{fR}\right)_{\min}$ are presented as follows (five significant figures retained):
\begin{equation}
\left\{ \begin{gathered}
  {{C_{fD}}^*\left(\alpha_{fD}\right)_{\min}} = 2.9851 \hfill \\
{{C_{fR}}^*\left(\alpha_{fR}\right)_{\min}} = 2.9420 \hfill \\ 
\end{gathered}  \right.
 \label{eq50}
\end{equation} 

\begin{remark}[1]
The most direct theoretical contribution made by the analytical energy conditions is that these conditions suggest an important role of the solar gravity perturbation in achieving lunar ballistic capture. According to initial guesses generated by Table 2 and initial Jacobi energy shown in Fig. 4 (a) in Ref. \cite{oshima2019low}, it implies that adopting initial guesses generated by Table 2 to generate ballistic lunar transfers in the Earth-Moon PCR3BP may be impossible ($C_i<2.8$ for the obtained solutions in Ref. \cite{oshima2019low}). In Ref. \cite{oshima2019low}, however, lunar ballistic capture is achieved (shown in Fig. 3 (b) in Ref. \cite{oshima2019low}). It is the solar gravity perturbation that make the Jacobi energy vary to satisfy the analytical conditions Theorems \ref{th3.1}-\ref{th3.3}.
\end{remark}

\begin{remark}[2]
According to Eq. \eqref{eq11}, the difference between the angular momentum with respect to the Moon for direct and retrograde insertion leads directly to the difference between ${C_{fD}}^*{\left( {{\alpha _{fD}}} \right)} $ and ${C_{fR}}^*{\left( {{\alpha _{fR}}} \right)} $. Although these values can also be found by solving $E_{fD}=0$ and $E_{fR}=0$, our derivation clearly reveals the cause of the difference. This derivation is also suitable for ballistic capture analysis in the Earth-Moon PCR3BP (where the Jacobi energy is constant) and agrees with the numerical findings obtained from Ano{\`e} et al. \cite{anoe2024ballistic}. Moreover, differing from Ref. \cite{anoe2024ballistic}, our derivation of ${C_{fD}}^*{\left( {{\alpha _{fD}}} \right)} $ and ${C_{fR}}^*{\left( {{\alpha _{fR}}} \right)} $ can be further extended to systems with different values of $\mu$ or to cases with different altitudes of circular orbits around the secondary body, without requiring large-scale computation.
\end{remark}

\begin{remark}[3]
Our derivation of ${C_{fD}}^*{\left( {{\alpha _{fD}}} \right)} $ and ${C_{fR}}^*{\left( {{\alpha _{fR}}} \right)} $ suggests a possible explanation for the bifurcation between stable and unstable Moon collision trajectories, as discovered by Oshima et al. \cite{oshima2017analysis,oshima2019linking}, from an energy perspective. Specifically, when $r_f$ is set to 0, we have ${C_{fD}}^*{\left( {{\alpha _{fD}}} \right)} ={C_{fR}}^*{\left( {{\alpha _{fR}}} \right)} =3\left(1-\mu\right)$, which agrees with the exact value of the Jacobi energy at the bifurcation point \cite{oshima2017analysis,oshima2019linking}. This is an interesting coincidence. In this case, when $C_f>3\left(1-\mu\right)$, the Moon collision trajectories are hyperbolic trajectories with respect to the Moon and exhibit instability.
\end{remark}

\begin{remark}[4]
In 2004, Belbruno \cite{belbruno2004capture} presented the approximation value of the Jacobi energy when the trajectory with respect to the Moon is a parabolic trajectory in the Earth-Moon PCR3BP. He presented this approximation value as $C_f^* \approx3$ under the assumptions $r_f \approx0$ and $\mu \approx 0$. However, to determine the construction parameters for ballistic lunar transfers, more exact ranges of the Jacobi energy at the insertion point are required than those provided by the previous approximation. Therefore, we derive the closed-form expressions of ${C_{fD}}^*\left(\alpha_{fD}\right)$ and ${C_{fR}}^*\left(\alpha_{fR}\right)$ without approximation.
\end{remark}

\begin{remark}[5]
In 2010, Fantino et al. \cite{fantino2010note} developed the closed-form expressions of the Jacobi energy as functions of $r_f$, $\alpha_f$, and eccentricity with respect to the Moon $e_{\text{M}}$. These expressions are equivalent to the expressions of ${C_{fD}}^*\left(\alpha_{fD}\right)$ and ${C_{fR}}^*\left(\alpha_{fR}\right)$ when $e_{\text{M}}$ is set to 1. However, they originally developed them only to calculate WSB points and did not analyze the monotonicity relationship between the Jacobi energy and $e_{\text{M}}$ ($e_{\text{M}}$ is equivalent to $E_f$). Although this monotonicity is not necessary for WSB calculation, it important for the analysis of the energy conditions. To our best knowledge, these conditions have not been comprehensively analyzed, and our work provides this theoretical supplementation. Meanwhile, we weaken the obtained sufficient and necessary condition to the necessary condition, further distinguishing our contributions from Ref. \cite{fantino2010note}.
\end{remark}

\begin{remark}[6]
The obtained analytical energy conditions can further provide insights into constructing ballistic lunar transfers. The obtained necessary condition Theorem \ref{th3.3} is used to determine feasible ranges of construction parameters, detailed in Section \ref{sec4}.
\end{remark}

Subsequently, the obtained energy conditions are applied to constructing ballistic lunar transfers.

\section{APPLICATION TO CONSTRUCTING BALLISTIC LUNAR TRANSFERS}\label{sec4}
This section proposes a grid-search method combined with analytical energy conditions to construct ballistic lunar transfers. In particular, the derived energy conditions Theorem \ref{th3.3} provides an analytical threshold for grid search and trajectory correction. Then, the results obtained from this method are discussed.
\subsection{Grid-Search Method Combined with Analytical Energy Conditions}\label{subsec4.1}
When constructing lunar transfers in the Sun-Earth/Moon PBCR4BP, the grid-search method is adopted. In particular, the method developed in this paper uses the analytical energy conditions to provide an analytical threshold for grid search and trajectory correction. Because the necessary conditions Theorem \ref{th3.3} focus on the Jacobi energy at the insertion point, we propose a backward strategy, i.e., we select the states at the insertion point, and perform backward time propagation to search the transfer trajectories satisfying the constraints Eqs. \eqref{eq5}-\eqref{eq6}. The specific procedure is summarized as follows, including performing grid search, generating initial guesses, and performing differential correction.
\subsubsection{Performing Grid Search}\label{subsubsec4.1.1}
The construction parameters are selected as $\alpha_f$, $C_f$, and $\theta_{\text{S}f}=\omega_{\text{S}}t_f$. To construct lunar transfers with direct capture, the search conditions are set as $\alpha_{fD}\in \left[0,\text{ }2\pi\right)$ with a step-size of $\pi/360$, $C_{fD} \in \left[2.9851,\text{ }3.2003\right]$ with a step-size of $0.0001$, and $\theta_{\text{S}fD} \in \left[0,\text{ }2\pi\right)$ with a step-size of $\pi/360$. Among these parameters, the lower boundary of $C_{fD}$ is selected as ${{C_{fD}}^*\left(\alpha_{fD}\right)_{\min}}$ (i.e., the analytical threshold) according to Theorem \ref{th3.3}. When determining the upper boundary of $C_{fD}$, we refer to the results shown in Fig. 4 (b) of Ref. \cite{oshima2019low} and select it as the Jacobi energy at the L1 libration point $C_{\text{L1}}=3.2003$. Similarly, the search conditions to construct lunar transfers with retrograde capture are set as $\alpha_{fR}\in \left[0,\text{ }2\pi\right)$ with a step-size of $\pi/360$, $C_{fR} \in \left[2.9420,\text{ }3.2003\right]$ with a step-size of $0.0001$, and $\theta_{\text{S}fR} \in \left[0,\text{ }2\pi\right)$ with a step-size of $\pi/360$. In the previous works, Topputo \cite{topputo2013optimal} and Oshima et al. \cite{oshima2019low} parameterized the grid in terms of the states at the departure point. This paper parameterizes the grid in terms of the states at the insertion point. The analytical threshold of $C_f$ is added based on the analytical energy conditions, yielding a complementary perspective.
\subsubsection{Generating Initial Guesses}\label{subsubsec4.1.2}
With the search conditions selected in Section IV-A-\ref{subsubsec4.1.1}, the states of insertion point can be calculated by Eqs. \eqref{eq13} or \eqref{eq14}. Then, we perform backward time propagation of these states, and the propagation time is set to 200 days. Because the states calculated by Eqs. \eqref{eq13} or \eqref{eq14} satisfy the constraint Eq. \eqref{eq6} rigorously, we focus on the residual of Eq. \eqref{eq5}. When the states during the propagation satisfy:
\begin{equation}
\left\| {{\bm{\psi} _i}} \right\| < 1 \times {10^{ - 4}}
\label{eq501}
\end{equation}
the corresponding construction parameters are recorded as the initial guesses of the states at the insertion point, and the epoch of sates satisfying Eq. \eqref{eq501} is recorded as an initial guess of the departure epoch $t_i$. Here we transform the initial guesses of $t_i$ into the initial guesses of time of flight ($\text{TOF}=\theta_{\text{S}f}/{{\omega_{\text{S}}}}-t_i$). Based on the aforementioned discussion, the parameters determining an initial guess trajectory can be expressed as:
\begin{equation}
\bm{y}=\left[\alpha_f,\text{ }C_f,\text{ }\theta_{\text{S}f},\text{ TOF}\right]^{\text{T}}
\label{eq502}
\end{equation}
With these parameters, initial guesses are generated using backward time propagation. Notably, the Earth/Moon collision trajectories \cite{topputo2013optimal,oshima2019low} are excluded during the generation of initial guesses.
\subsubsection{Performing Differential Correction}\label{subsubsec4.1.3}
When performing differential correction, the problem is transformed into a nonlinear programming (NLP) problem \cite{topputo2013optimal,liu2026survey}. The NLP variables are set as $\bm{y}$ defined in Eq. \eqref{eq502}, and the constraints of the NLP problem are set as Eq. \eqref{eq5}, and the function to minimize is set as 0 (here the NLP problem is equivalent to a shooting problem). In simulations, the NLP problem is solved by MATLAB®'s fmincon command with the sequential quadratic programming method. The parameters of the fmincon command are selected through trial and error to ensure efficiency and accuracy, as presented in Table \ref{tab2}.
\begin{table}[!htb]
\caption{Parameter settings for fmincon command.}\label{tab2}%
\centering
\renewcommand{\arraystretch}{1.5}
\begin{tabular}{@{}ll@{}}
\hline
Parameter & Value   \\
\hline
StepTolerance    & $1 \times {10^{ - 13}}$       \\
FunctionTolerance    & $1 \times {10^{ - 8}}$       \\
ConstraintTolerance    & $1 \times {10^{ - 8}}$      \\
MaxIterations    & $1000$    \\
MaxFunctionEvaluations    & $1000$      \\
\hline
\end{tabular}
\end{table}

The lower and upper boundaries of the NLP variables are presented in Table \ref{tab3} ($T_0$ denotes the dimensionless time of 200 days and $\theta_{\text{S}f}$ is selected as the value of initial guess parameter). Notably, the lower boundaries of $C_f$ are selected based on the necessary conditions Theorem \ref{th3.3}.
\begin{table*}[h]
\centering
\renewcommand{\arraystretch}{1.5}
\caption{Lower and upper boundaries of the NLP variables.}\label{tab3}%
\begin{tabular}{@{}lll@{}}
\hline
NLP Variables    & Lower Boundary & Upper Boundary  \\
\hline
$\alpha_{fD}$, $\alpha_{fR}$ & $-\text{Inf}$, $-\text{Inf}$ & $\text{Inf}$, $\text{Inf}$ \\
$C_{fD}$, $C_{fR}$ & $2.9851$, $2.9420$ & $3.2003$, $3.2003$ \\
$\theta_{\text{S}fD}$, $\theta_{\text{S}fR}$ & $\theta_{\text{S}fD}-\pi$, $\theta_{\text{S}fR}-\pi$ & $\theta_{\text{S}fD}+\pi$, $\theta_{\text{S}fR}+\pi$ \\
$t_{iD}$, $t_{iR}$ & $\theta_{\text{S}fD}/\omega_{\text{S}}-T_0$, $\theta_{\text{S}fR}/\omega_{\text{S}}-T_0$ &$\theta_{\text{S}fD}/\omega_{\text{S}}-\pi/10$, $\theta_{\text{S}fR}/\omega_{\text{S}}-\pi/10$ \\
\hline
\end{tabular}
\end{table*}

After differential correction, lunar transfers satisfying the tolerance of the constraints (i.e., $\left\| {{\bm{\psi} _i}} \right\| < {1 \times {10^{ - 7}}}$) are recorded. For practical consideration, only the lunar transfers from the prograde Earth parking orbit are recorded. During the correction, the Earth/Moon collision trajectories are also excluded. 

\begin{remark}[7]
The solutions satisfying Eqs. \eqref{eq5}-\eqref{eq6} can be obtained from the aforementioned procedure. To obtain more feasible solutions and improve the convergence in the correction process, the continuation method \cite{topputo2013optimal,oshima2019low,liu2024design,fu2025analytical} and multiple shooting method \cite{topputo2013optimal,oshima2019low} can be used.
\end{remark}

Subsequently, the results are presented and analyzed.

\subsection{Results and Discussion}\label{subsec4.2}
\begin{figure*}
\centerline{\includegraphics[width=25pc]{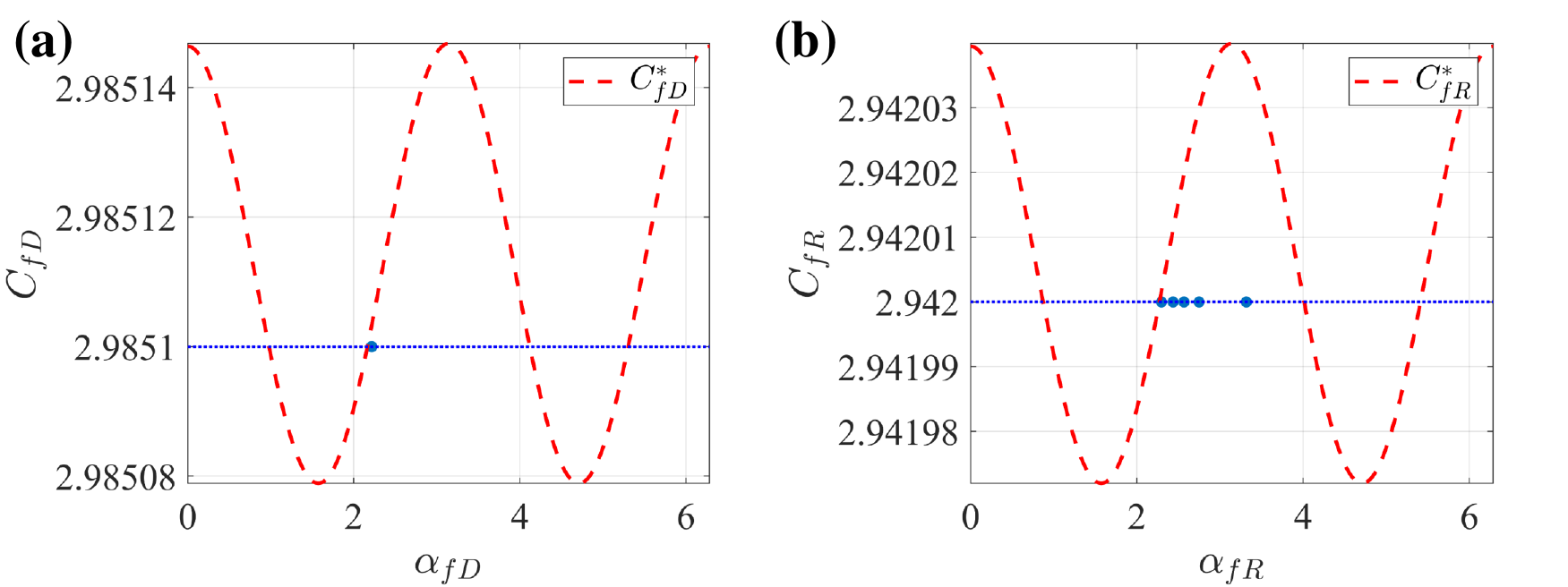}}
\caption{The values of $\alpha_{f}$ and $C_{f}$ of the transfer trajectories without ballistic capture. (a) Direct insertion; (b) Retrograde insertion.}
\label{fig_hyperbolic}
\end{figure*}

After correction, we obtain 1982 solutions departing from the prograde Earth parking orbit, including 1591 solutions with direct insertion and 391 solutions with retrograde insertion. The construction parameters of the obtained solutions can be found in the supplementary materials. Investigating the values of $E_f$ of these solutions, the ratio of ballistic capture is $1589/1591$ (99.87$\%$) for direct insertion and $386/391$ (98.72$\%$) for retrograde insertion, respectively. The values of $\alpha_{f}$ and $C_{f}$ of the transfer trajectories without ballistic capture for direct and retrograde insertion is shown in Fig. \ref{fig_hyperbolic}. The red curves show the variation of ${C_{fD}}^*$ (${C_{fR}}^*$) with respect to $\alpha_{f}$, and the blue lines show the values of the selected lower boundaries of $C_{f}$ in Table \ref{tab3}. It is observed that $C_{f}$ of this trajectory is higher than the selected lower boundaries but lower than ${{C_{fD}}^*}$ (${{C_{fR}}^*}$), explaining why it does not achieve ballistic capture. In the following texts, we focus on the obtained ballistic lunar transfers. 

\begin{figure*}
\centerline{\includegraphics[width=25pc]{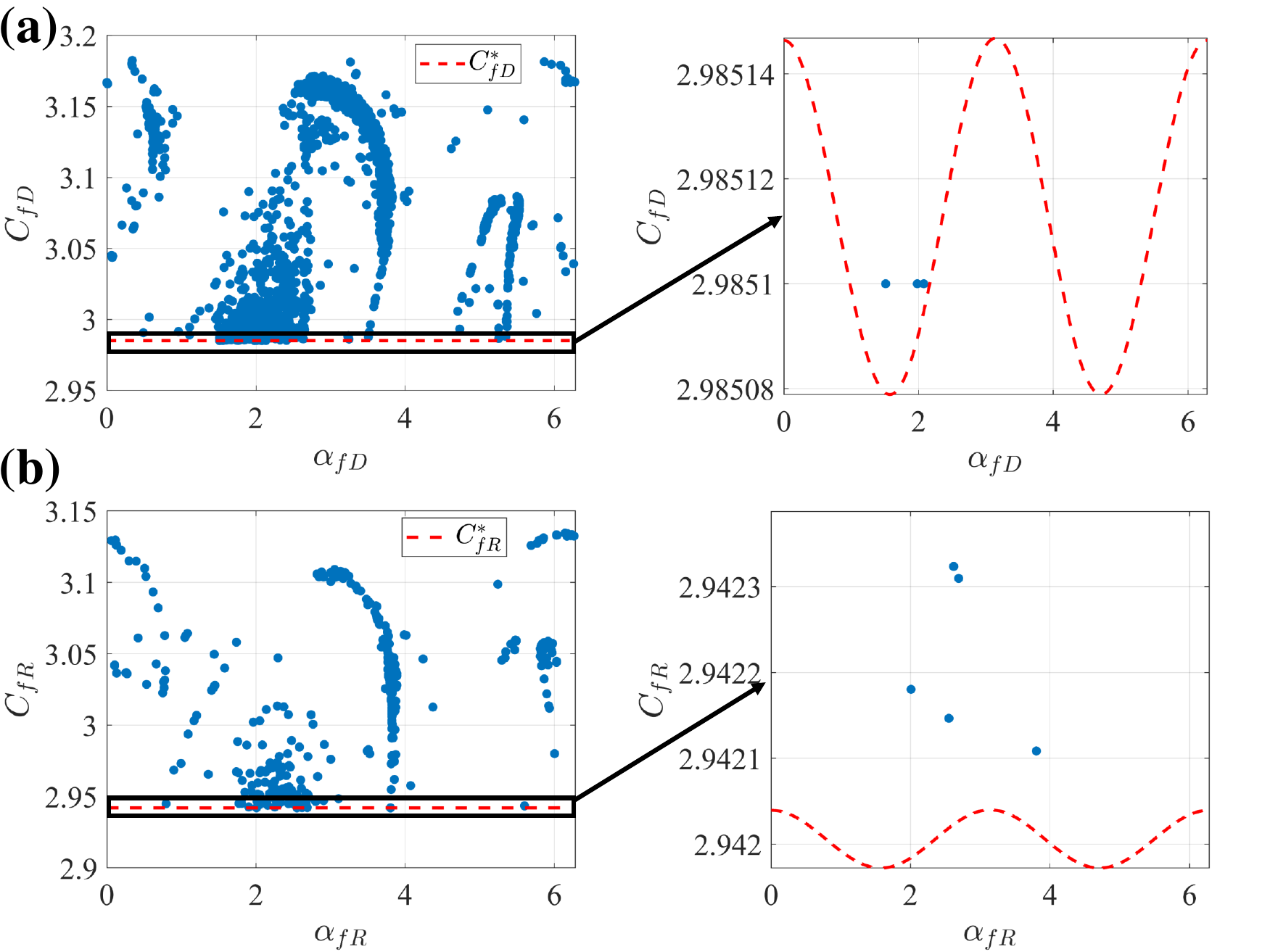}}
\caption{The distributions of $C_f$ with respect to $\alpha_f$ of the obtained transfers with direct and retrograde capture. (a) Direct capture; (b) Retrograde capture.}
\label{fig_CfD_CfR}
\end{figure*}

The distributions of $C_f$ with respect to $\alpha_f$ of the obtained transfers with direct and retrograde capture are shown in Fig. \ref{fig_CfD_CfR}. From this figure we can observe that all of the values of $C_f$ satisfy $C_f\left(\alpha_{f}\right)\geq {C_{f}}^*\left(\alpha_{f}\right)$ for direct and retrograde capture, which illustrates that ${C_{f}}^*\left(\alpha_{f}\right)\leq C_f\left(\alpha_{f}\right)\leq W\left(\alpha_{f}\right)$ ($W\left(\alpha_{f}\right)\approx 8.0486$). Therefore, the numerical results of the trajectories with ballistic capture are consistent with the derived analytical conditions, including both the necessary and sufficient conditions Theorem \ref{th3.1} and necessary conditions Theorem \ref{th3.3}. Subsequently, the transfer characterizations of these ballistic lunar transfers are analyzed.

\begin{figure*}
\centerline{\includegraphics[width=25pc]{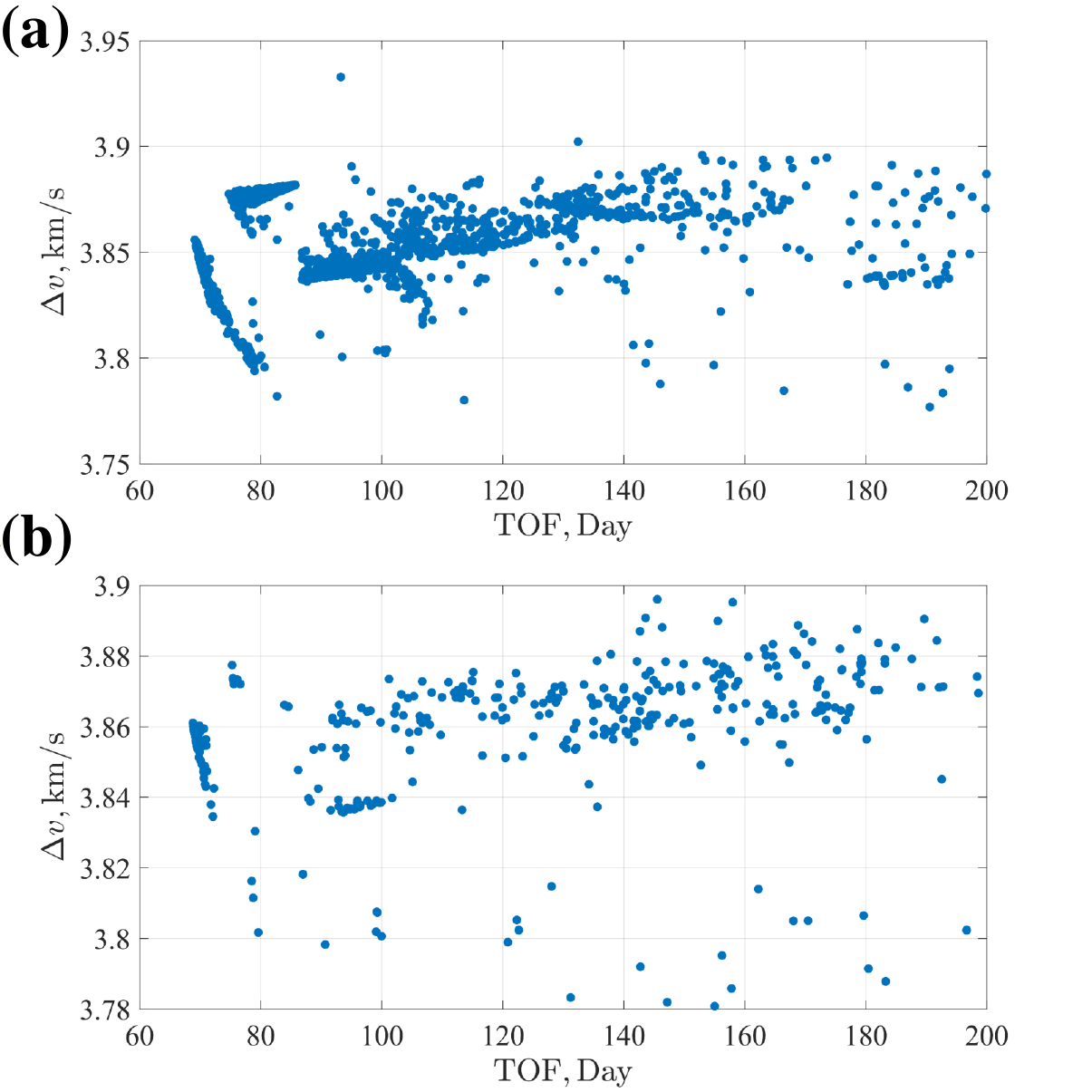}}
\caption{The $\left(\text{TOF},\text{ }\Delta{v}\right)$ maps. (a) Lunar transfers with direct capture; (b) Lunar transfers with retrograde capture.}
\label{fig_DV}
\end{figure*}

Figure \ref{fig_DV} presents the $\left(\text{TOF},\text{ }\Delta{v}\right)$ maps of the obtained ballistic lunar transfers. Here we select the typical samples with direct and retrograde capture, and provide detailed information about these transfers.

\subsubsection{Direct Capture: Samples I-III}\label{subsubsec4.2.1}
\begin{figure*}
\centering
\includegraphics[width=40pc]{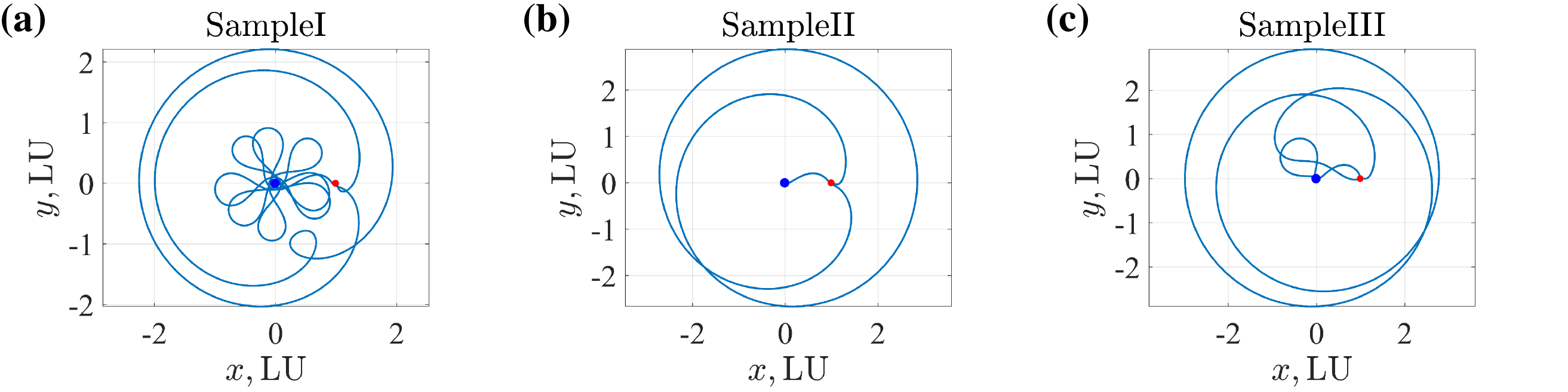}
\caption{Typical samples with direct capture. (a) Sample I; (b) Sample II; (c) Sample III.}
\label{fig_trajectory}
\end{figure*}

\begin{figure*}
\centerline{\includegraphics[width=25pc]{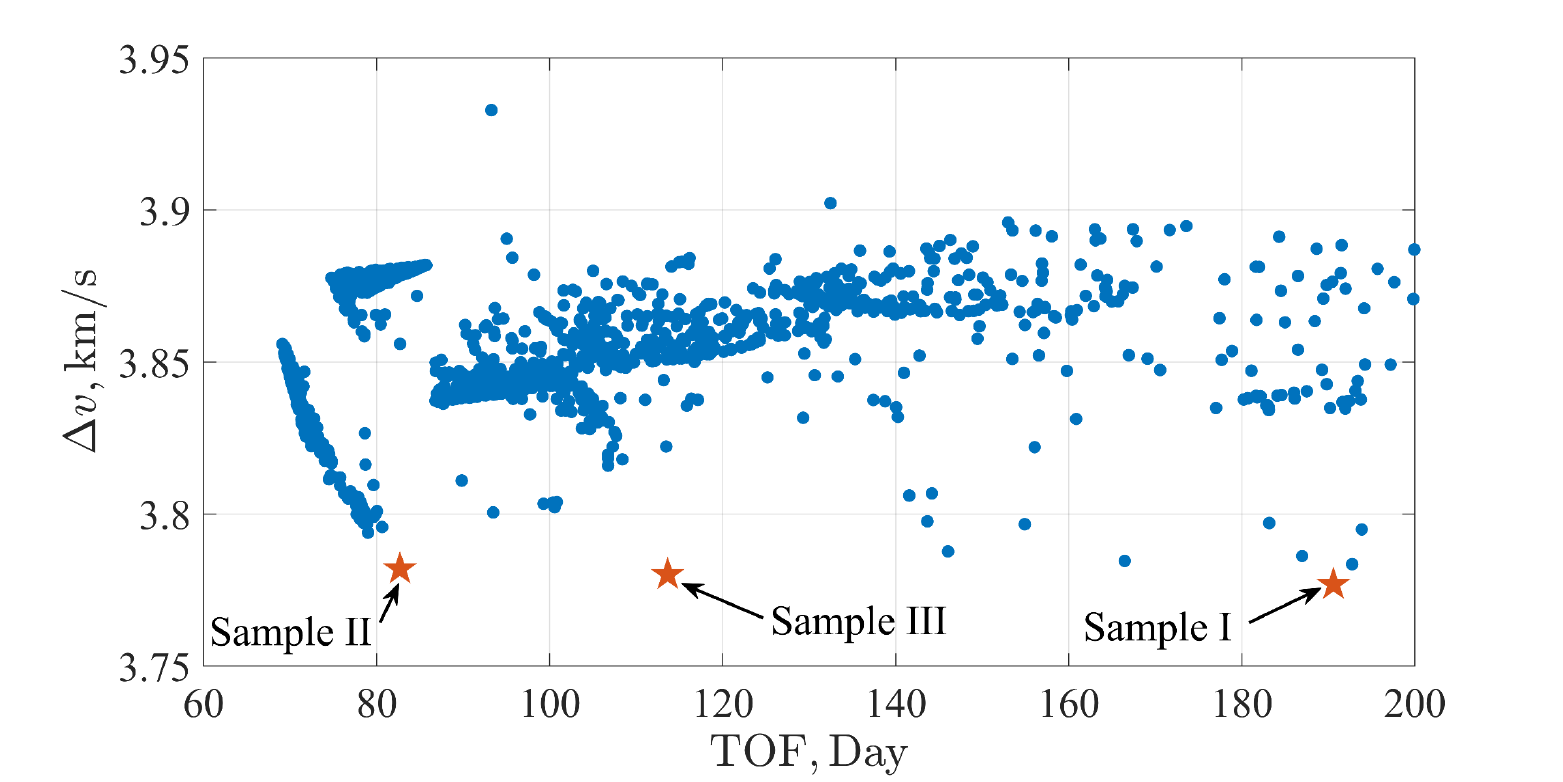}}
\caption{The $\left(\text{TOF},\text{ }\Delta{v}\right)$ maps (direct capture). The red stars denote the selected samples.}
\label{fig_DV_DC_Low_DV}
\end{figure*}

The typical trajectories with direct samples are presented in Fig. \ref{fig_trajectory}. From Fig. \ref{fig_trajectory}, it is observed that all of the three samples are exterior transfers in the Sun-Earth/Moon PBCR4BP. Samples I and II involve a direct LGA, while Sample III involves a retrograde LGA (the LGA can be categorized into direct and retrograde LGA according to the sign of the angular momentum with respect to the Moon \cite{yang2025deep}). Moreover, Sample I further belongs to ‘interior-exterior’ transfers (which can be considered as new transfer families discovered by Oshima et al. \cite{oshima2019low}) and involves a LGA. It performs several revolutions around the Earth and finally encounters the Moon through the L2 region. This type of transfer takes advantage of LGA and high-altitude lunar flyby to reduce $\Delta{v}$. The corresponding $\left(\text{TOF},\text{ }\Delta{v}\right)$ of these three samples are shown in Fig. \ref{fig_DV_DC_Low_DV}. These three samples correspond to the three lowest $\Delta{v}$ values with direct capture, and the minimum $\Delta{v}$ belongs to Sample I. The minimum $\Delta{v}$ of the obtained solutions with direct capture is $3.777\text{ km/s}$.

\subsubsection{Retrograde Capture: Samples IV-VI}\label{subsubsec4.2.2}

\begin{figure*}
\centering
\includegraphics[width=40pc]{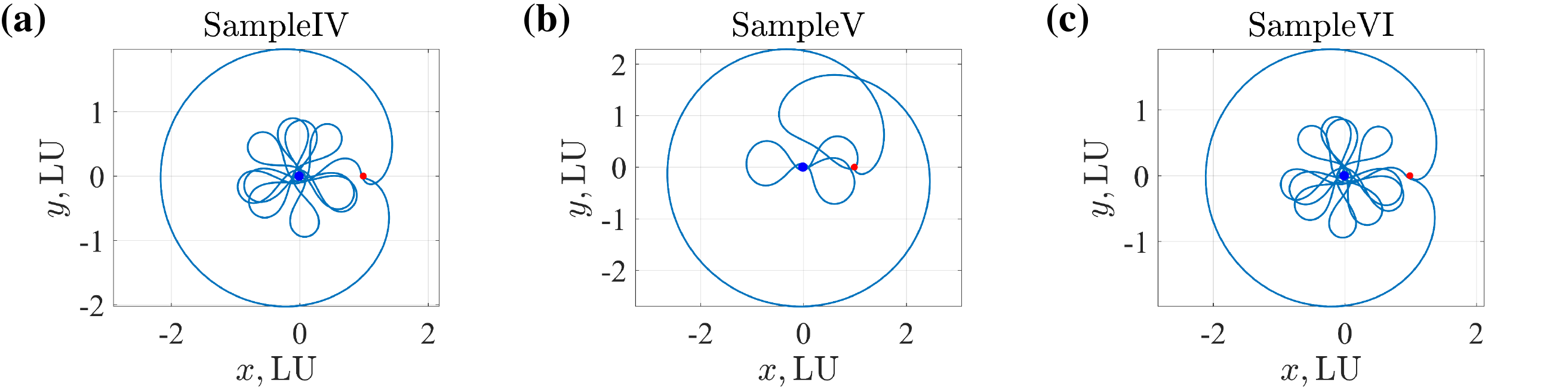}
\caption{Typical samples with retrograde capture. (a) Sample IV; (b) Sample V; (c) Sample VI.}
\label{fig_trajectory_RC}
\end{figure*}

\begin{figure*}
\centerline{\includegraphics[width=25pc]{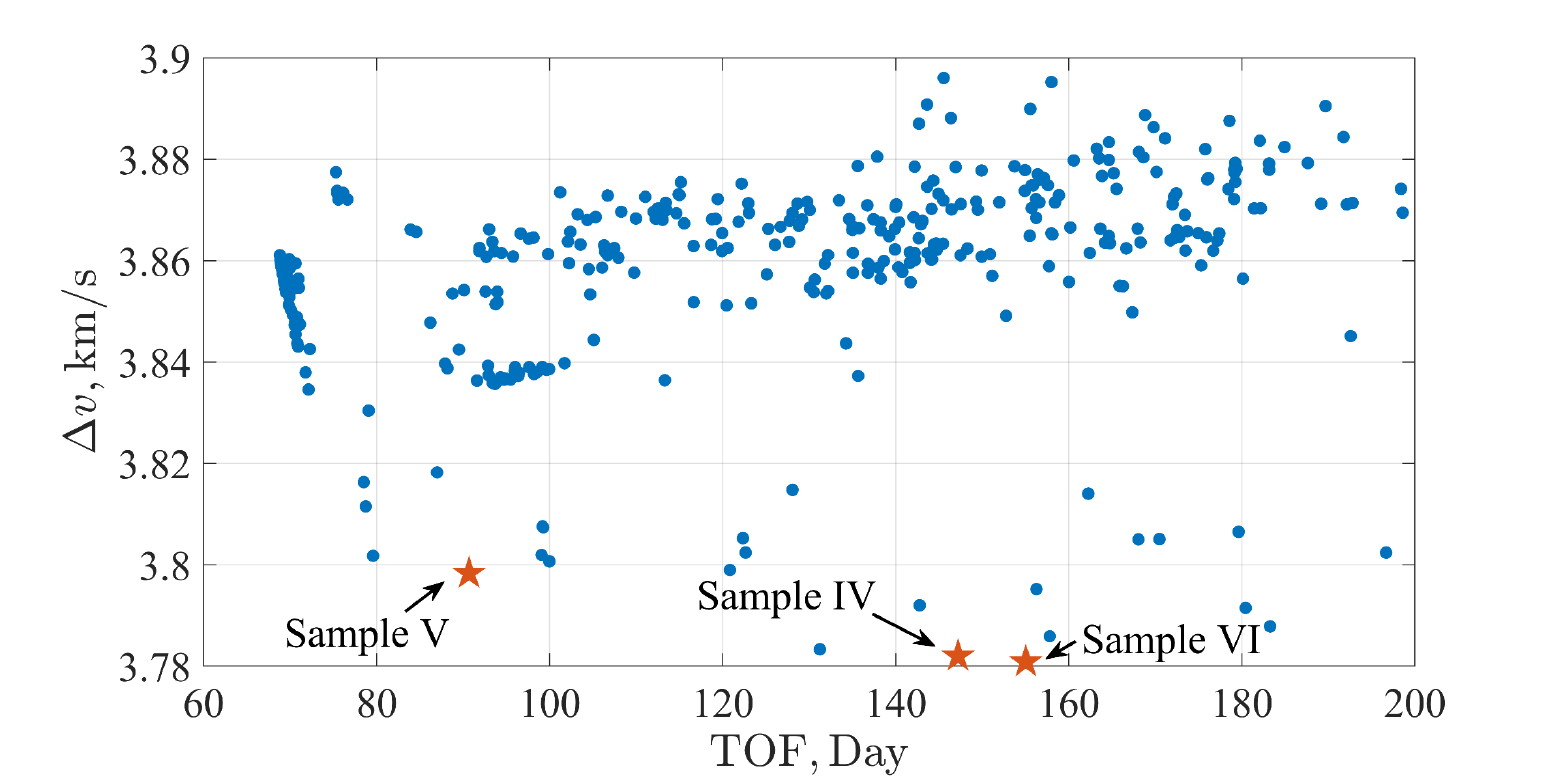}}
\caption{The $\left(\text{TOF},\text{ }\Delta{v}\right)$ maps (retrograde capture). The red stars denote the selected samples.}
\label{fig_DV_RC_Low_DV}
\end{figure*}

The typical trajectories with direct samples are presented in Fig. \ref{fig_trajectory_RC}. From Fig. \ref{fig_trajectory_RC}, it is observed that all of the three samples can be considered as ‘interior-exterior’ transfers in the Sun-Earth/Moon PBCR4BP. Among these three samples, Samples IV and VI involve a direct LGA while Sample V involves a retrograde LGA. The corresponding $\left(\text{TOF},\text{ }\Delta{v}\right)$ of these three samples are shown in Fig. \ref{fig_DV_RC_Low_DV}. The minimum $\Delta{v}$ for the retrograde capture case belongs to Sample VI, and its value is $3.781\text{ km/s}$. Moreover, when checking the obtained solutions with retrograde capture, we find some special transfer families which are new or less-reported compared to the previous works \cite{topputo2013optimal,oshima2019low,oshima2017analysis,qi2017optimal,campana2024clustering,campana2025ephemeris}. The corresponding samples are presented in the following texts.

\subsubsection{Retrograde Capture: Samples VII-IX}\label{subsubsec4.2.3}

\begin{figure*}
\centering
\includegraphics[width=40pc]{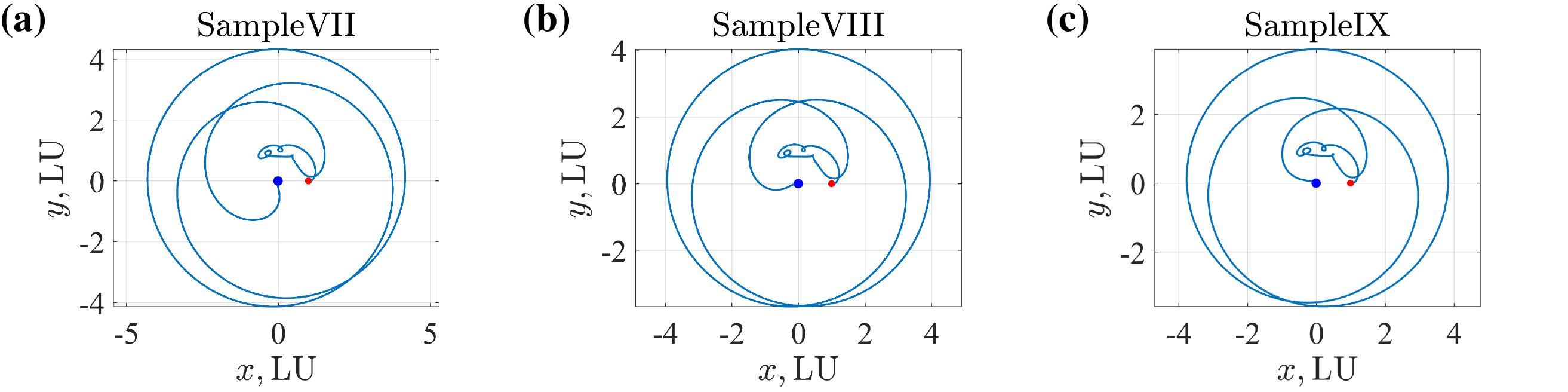}
\caption{Typical samples with retrograde capture. (a) Sample VII; (b) Sample VIII; (c) Sample IX.}
\label{fig_trajectory_RC_New_Family}
\end{figure*}

\begin{figure*}
\centerline{\includegraphics[width=25pc]{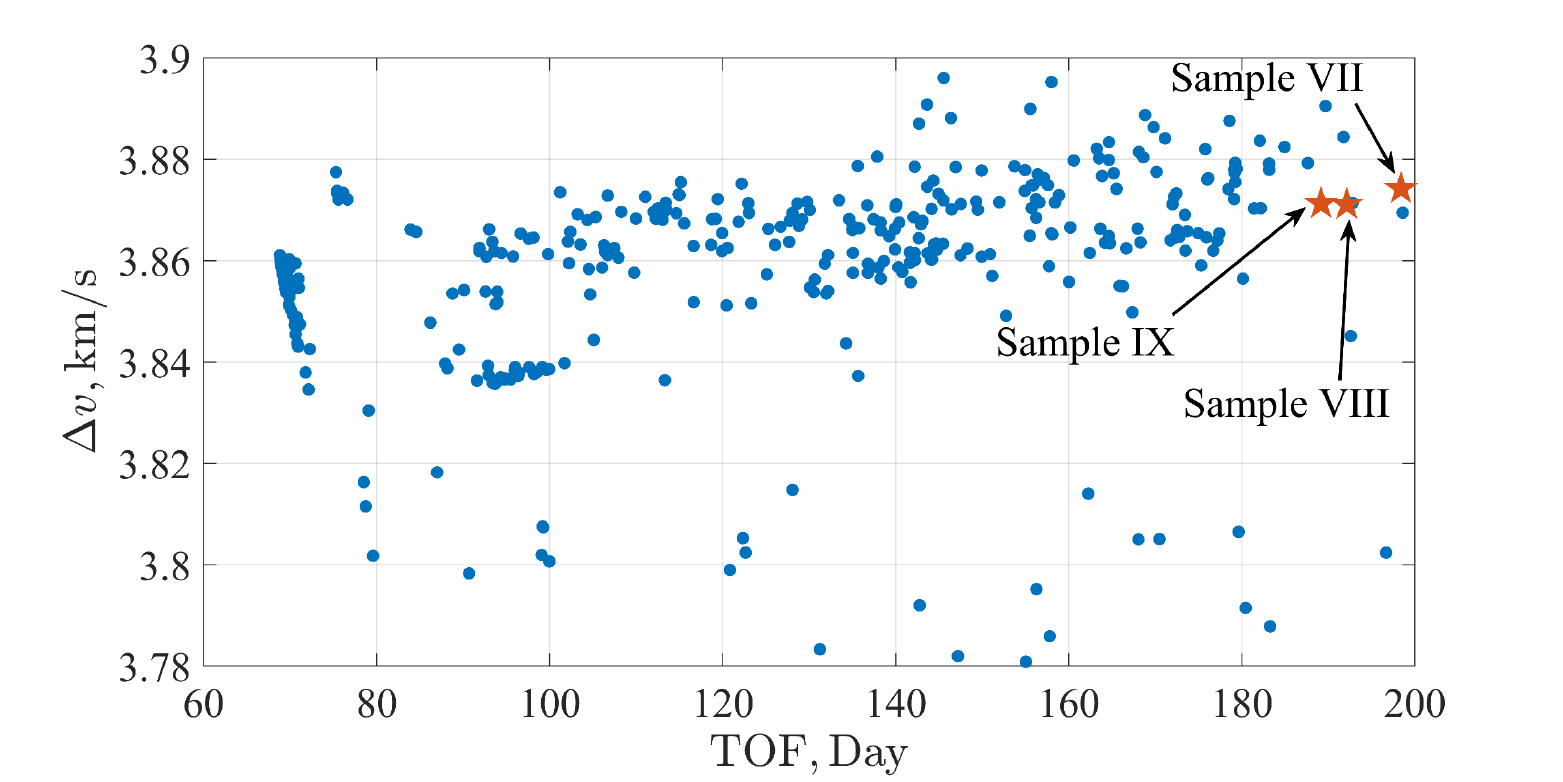}}
\caption{The $\left(\text{TOF},\text{ }\Delta{v}\right)$ maps (retrograde capture). The red stars denote the selected samples.}
\label{fig_DV_RC_New_Family}
\end{figure*}

The samples belonging to new or less-reported transfer families are presented in Fig. \ref{fig_trajectory_RC_New_Family}. These three samples are also exterior transfers. However, it is worth to note that before achieving ballistic capture, the trajectories move in tadpole-like orbits \cite{oshima2015jumping} in the L4 region. This is an interesting phenomenon. These trajectories may be suitable for the single-launch dual-mission framework: one mission to explore the Moon, and the other one mission to explore the Earth-Moon L4 region. An illustration example for preliminarily design can be found in Appendix. The corresponding $\left(\text{TOF},\text{ }\Delta{v}\right)$ of these three samples are shown in Fig. \ref{fig_DV_RC_New_Family}. It can be found that although the total impulses $\Delta{v}$ of these three samples are relatively high compared to the aforementioned six samples, the trajectories can also achieve ballistic capture and insert into the circular lunar insertion orbit with a relatively low $\Delta v_f$. 
A reduced $\Delta v_f$ can allow for an increased payload mass of the spacecraft \cite{FU20254993}. Therefore, these new (or less-reported) trajectories still have remarkable engineering potentials.

\subsubsection{Link with Previous Works}\label{subsubsec4.2.4}

To further link our obtained solutions with those in previous works, we select conventional methods and typical solutions from previous works \cite{belbruno1993sun,Mingotti2012,Onozaki2017,oshima2017analysis,qi2017optimal,topputo2013optimal,oshima2019low} for comparison. To ensure comparability, we select the results satisfying the same constraints, i.e., the $\Delta v$ and TOF are calculated for trajectories departing a 167 km circular Earth parking orbit and inserting into a 100 km circular Moon insertion orbit. The comparison results are shown in Table \ref{tab5}. Among these, the methods from Refs. \cite{belbruno1993sun,Mingotti2012,Onozaki2017,oshima2017analysis,qi2017optimal} can be considered as methods only using prior knowledge about multi-body dynamics because they typically used dynamical structures (e.g., WSB, invariant manifolds, LCSs, dynamical structures of Moon collision trajectories and double-LGA trajectories) to generate initial guesses rather than using grid search. These methods typically yield fewer solutions and possibly miss the solutions with lower $\Delta{v}$. Our proposed method combines prior knowledge with grid search, which can explore the solution space more broadly and may find more solutions, including ones with lower $\Delta{v}$. From Table \ref{tab5}, it can be observed that Sample I of our obtained solutions yields a lower $\Delta{v}$ compared to most solutions obtained from methods only using prior knowledge \cite{belbruno1993sun,Mingotti2012,Onozaki2017,oshima2017analysis}. In particular, its $\Delta{v}$ is 61 m/s lower than the conventional WSB method \cite{belbruno1993sun}, and 16 m/s lower than the patched invariant manifolds method \cite{Mingotti2012}. Moreover, the value of $\Delta{v}$ of Sample I is comparable to the solution with minimum $\Delta{v}$ in Ref. \cite{qi2017optimal}. There also exist some solutions whose $\Delta{v}$ are lower than or comparable to solutions obtained from methods only using prior knowledge \cite{belbruno1993sun,Mingotti2012,Onozaki2017,oshima2017analysis,qi2017optimal} but result in shorter TOF compared to Sample I, such as Sample II ($\Delta{v}=3.782\text{ km/s}$, $\text{TOF}=83\text{ Day}$) and Sample V ($\Delta{v}=3.798\text{ km/s}$, $\text{TOF}=91\text{ Day}$). In particular, Sample II save 53 m/s of $\Delta{v}$ and 77 Day of TOF over solutions obtained from the WSB method. Moreover, the values of $\Delta{v}$ of samples obtained in this paper are slight higher than solutions with minimum $\Delta{v}$ in Refs. \cite{topputo2013optimal,oshima2019low}, as they explored the solution space of bi-impulsive lunar transfer more broadly using grid search combined with continuation methods. As a result, their results suggest a broader set of solutions, both with and without ballistic capture. With smaller step-sizes, multiple shooting, and continuation, solutions with lower $\Delta{v}$ similar to those in Refs. \cite{topputo2013optimal,oshima2019low} may be also found using our method.
\begin{table*}[h]
\centering
\renewcommand{\arraystretch}{1.5}
\caption{Results of the obtained solutions in this paper and previous works.}\label{tab5}%
\begin{tabular}{@{}lllll@{}}
\hline
Solution    & $\Delta{v},\text{ }\text{km/s}$  & TOF, Day & Model & Method \\
\hline
I & $3.777$  & 191 &  PBCR4BP & Grid search-Ballistic capture \\
II & $3.782$  & 83 &  &\\
III & $3.780$  & 114 &  &\\
IV & $3.782$  & 147 &  &\\
V & $3.798$  & 91 &  &\\
VI & $3.781$  & 155 &  &\\
VII & $3.874$  & 198 &  &\\
VIII & $3.871$  & 192 &  &\\
IX & $3.871$  & 189 &  &\\
\cite{belbruno1993sun} & $3.838$  & $160$ &  Ephemeris & WSB \\
\cite{Mingotti2012} & $3.793$ & $88$ &  PBCR4BP & Patched invariant manifolds \\
\cite{Onozaki2017} & $3.880$  & $100$ &  PBCR4BP & Patched LCSs \\
\cite{oshima2017analysis} & $3.820$ & $66$  &  PBCR4BP & Method using Moon collision trajectories\\
  & $3.859$ & $66$ &   &\\
\cite{qi2017optimal} & $3.777$ & $84$ & PBCR4BP & Method using double LGA trajectories\\
  & $3.802$ & $67$  & &\\
\cite{topputo2013optimal} & $3.769$ & $83$ &  PBCR4BP & Grid search-Continuation \\
\cite{oshima2019low} & $3.753$ & $183$  &  PBCR4BP & Grid search-Continuation\\
\hline
\end{tabular}
\end{table*}

\section{CONCLUSION}\label{sec5}
This paper derives the analytical energy conditions for lunar ballistic capture and proposes an grid-search method combined with these conditions to construct ballistic lunar transfers. The Sun-Earth/Moon planar bicircular restricted four-body problem is adopted to construct lunar transfers. First, analytical energy conditions for ballistic capture are derived. These conditions are expressed in the form of the specific ranges of the Jacobi energy at the insertion point of the lunar transfers. The important role of the solar gravity perturbation in achieving lunar ballistic capture is revealed by these conditions. Then, based on these conditions, a grid-search method is proposed and a backward strategy is employed to construct ballistic lunar transfers. Analytical conditions provide an analytical threshold for grid search and trajectory correction. For solutions obtained from our method, a ratio of ballistic capture is $1589/1591$ (99.87$\%$) for direct insertion and $386/391$ (98.72$\%$) for retrograde insertion. The values of the Jacobi energy at the insertion point of our obtained solutions with ballistic capture verify the effectiveness of the derived analytical energy conditions. Several samples are selected to perform a comparison with the solutions obtained from the conventional methods with the same constraints. The minimum impulse solution obtained in this paper is 3.777 km/s, which is lower than or comparable to most of the solutions obtained from the conventional methods. In particular, our obtained minimum impulse is 61 m/s lower than the weak stability boundary method and 16 m/s lower than the patched invariant manifolds method, respectively. New or less-reported transfer trajectories are found and provide a potential for designing a single-launch dual-mission framework: one mission for lunar exploration, and the other one mission for the Earth-Moon L4 exploration. The analytical energy conditions, the corresponding construction method, and obtained solutions provide a complementary perspective to the conventional model selection (the Sun-Earth/Moon PBCR4BP or the Earth-Moon PCR3BP) and grid-search method, establishing a direct link with the prior knowledge, trajectory construction, and mission design.

\section*{APPENDIX}
\begin{figure*}
\centerline{\includegraphics[width=25pc]{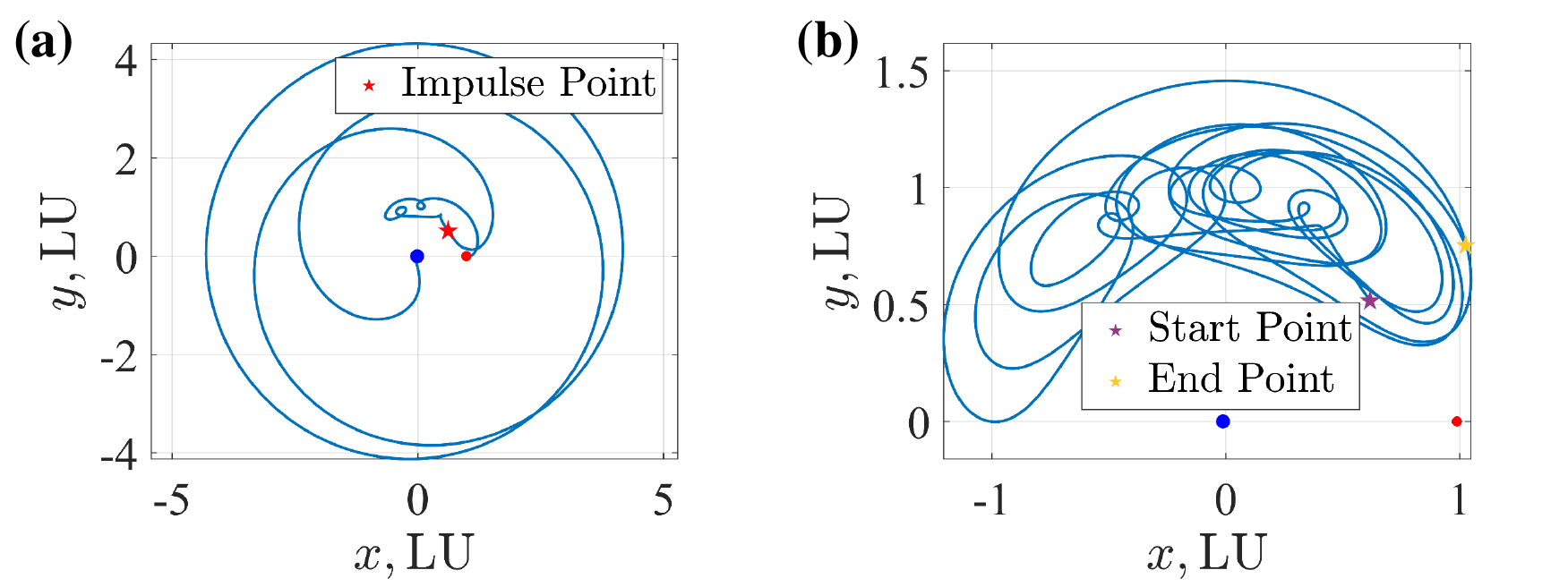}}
\caption{The $\left(\text{TOF},\text{ }\Delta{v}\right)$ maps (retrograde capture). The red stars denote the selected samples.}
\label{fig_DV_RC_New_Family_L4}
\end{figure*}

Figure \ref{fig_DV_RC_New_Family_L4} presents an example for preliminarily design of the application of trajectories reported in Section IV-B-\ref{subsubsec4.2.3} to a single-launch dual-mission framework: one mission for lunar exploration, and the other one mission for the Earth-Moon L4 exploration. The obtained lunar transfer trajectory is presented in Fig. \ref{fig_DV_RC_New_Family_L4} (a) (i.e., Sample VII), and the preliminarily design of the Earth-Moon L4 exploration mission is presented in Fig. \ref{fig_DV_RC_New_Family_L4} (b). Following injection into the trajectory shown in Fig. \ref{fig_DV_RC_New_Family_L4} (a), performing a 12 m/s impulse along the velocity tangential direction when 23.7 TU remain before insertion into the Moon insertion orbit will place the spacecraft into a large-amplitude tadpole-type trajectory in the Earth–Moon L4 region, as presented in Fig. \ref{fig_DV_RC_New_Family_L4} (b). The corresponding start and end points are marked with stars, with a TOF of 110 TU ($\approx 478 \text{ Day}$). This example may provide some supports for the potential engineering applications of the new (or less-reported) transfer trajectories mentioned in Section IV-B-\ref{subsubsec4.2.3}.

\bibliographystyle{IEEEtran}
\bibliography{sample}

\begin{IEEEbiography}[{\includegraphics[width=1in,height=1.25in,clip,keepaspectratio]{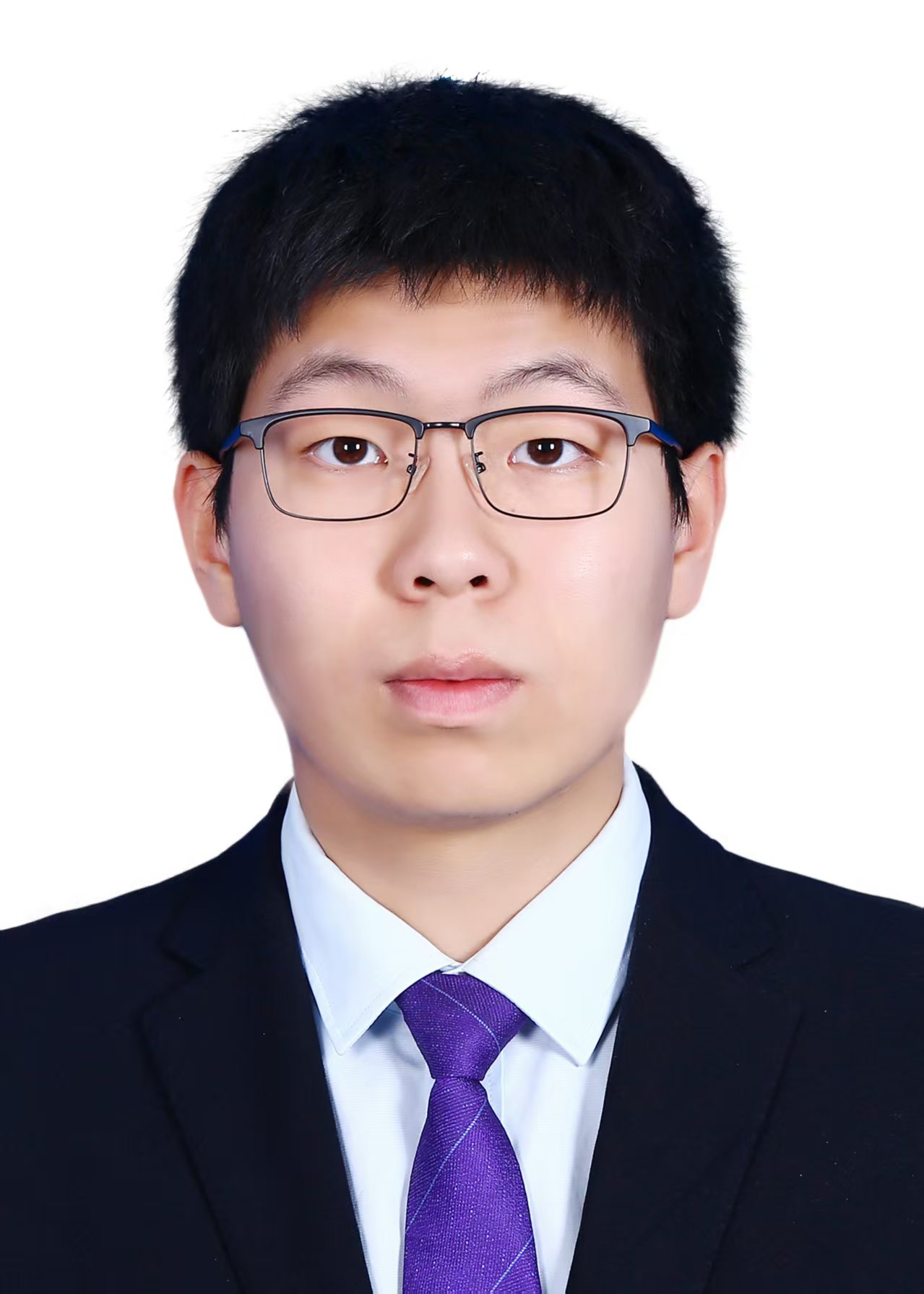}}]{Shuyue Fu} 
	received his B.E. degree in engineering in 2023 from Beihang University, Beijing, China, and
	he is currently working toward the Ph.D degree in astrodynamics and control with the School of Astronautics and Shen Yuan Honors College.
	His research interests include multi-body escape and capture dynamics, low-energy transfer, and orbital game.  
\end{IEEEbiography}

\begin{IEEEbiography}[{\includegraphics[width=1in,height=1.25in,clip,keepaspectratio]{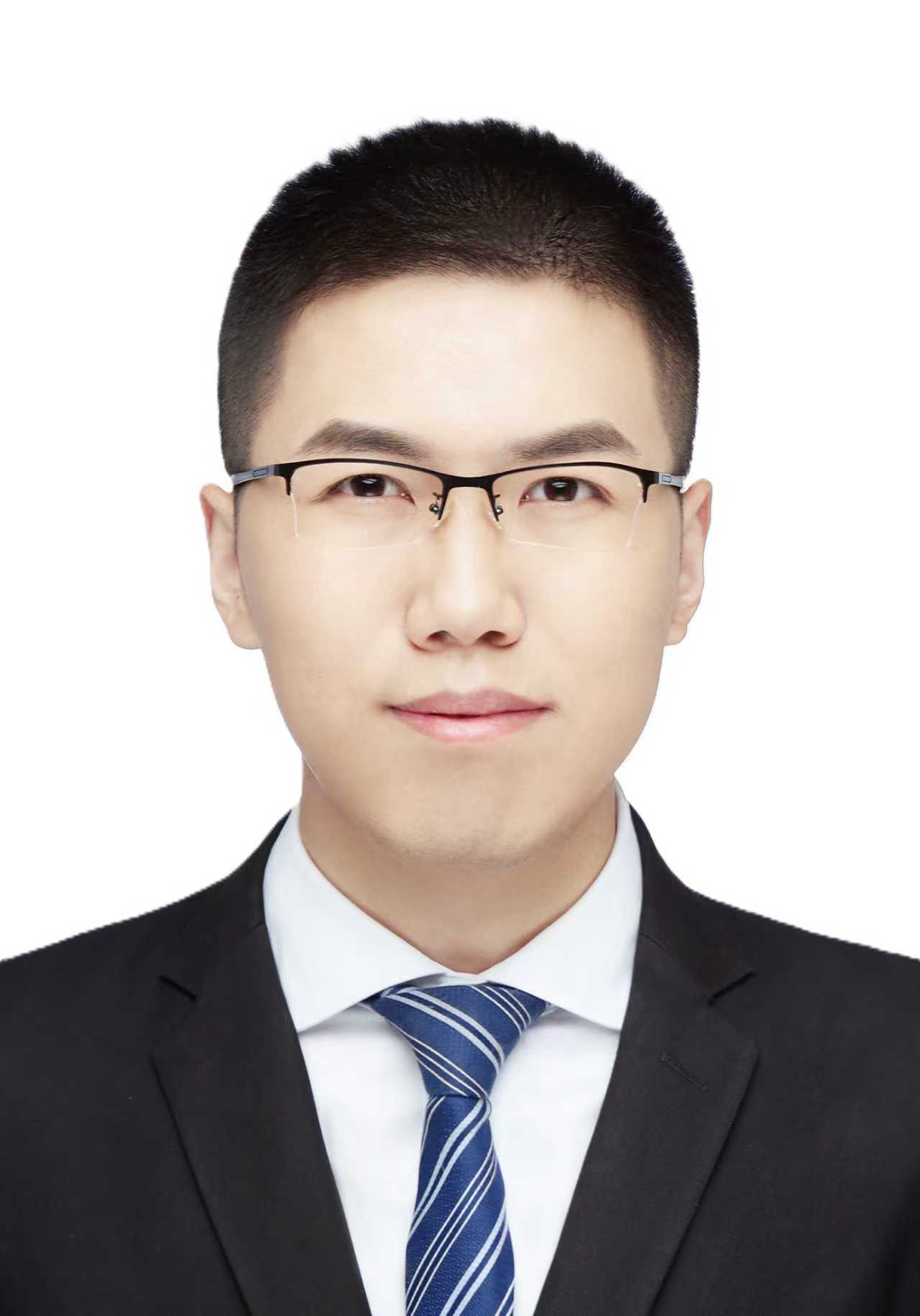}}]{Di Wu} received his Ph.D. degree, in 2022, from Tsinghua University. He is an associate professor at Beihang University, and his reasearch insterests include low-thrust trajectory optimization and mission design.
\end{IEEEbiography}

\begin{IEEEbiography}[{\includegraphics[width=1in,height=1.25in,clip,keepaspectratio]{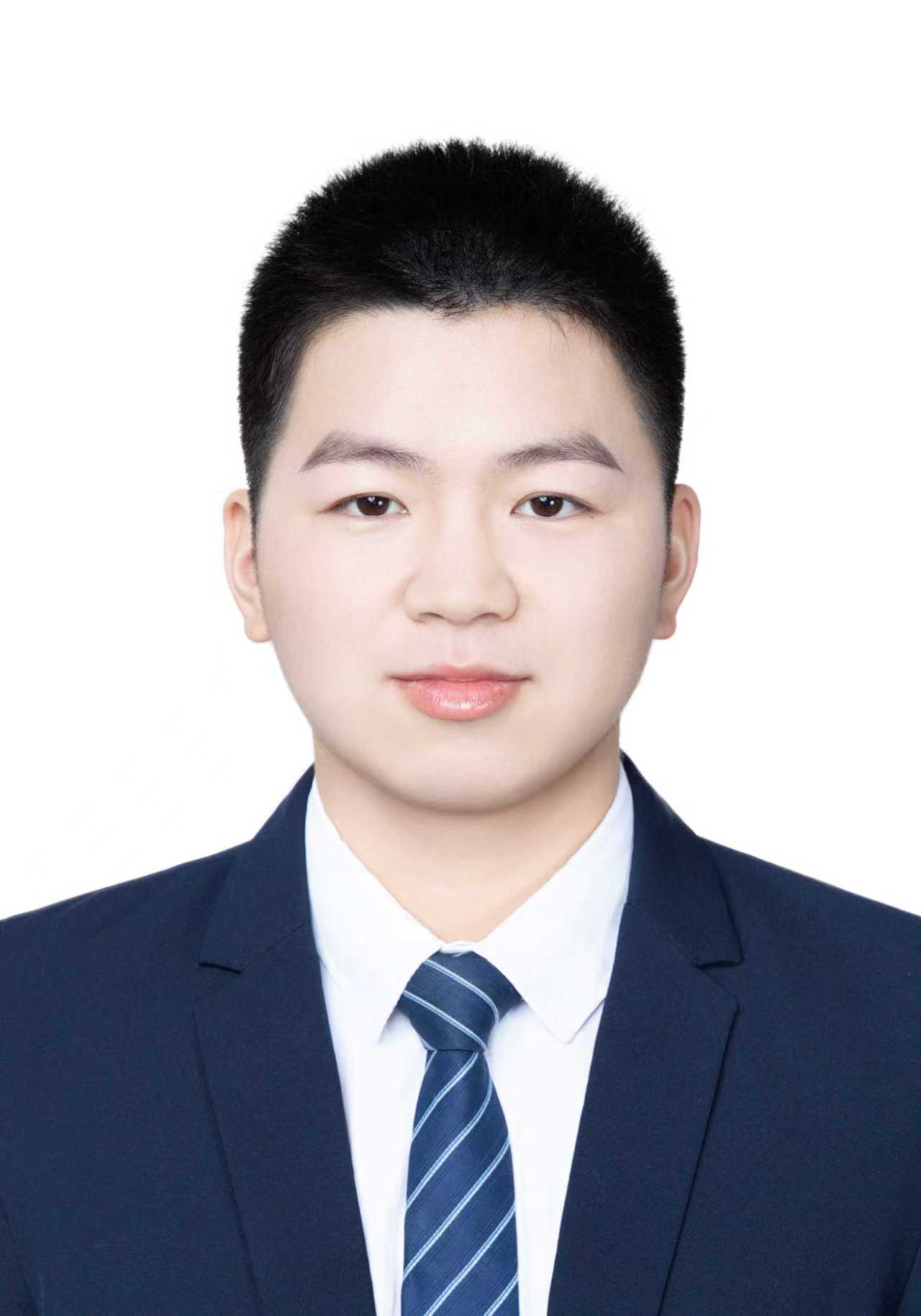}}]{Xiaowen Liu} 
	received his B.E. degree in engineering in 2024 from Beihang University, Beijing, China, and
	he is currently working toward the Ph.D degree in astrodynamics and control with the School of Astronautics.
	His research interests include optimal control, multi-body escape and capture dynamics, and low-energy transfer.  
\end{IEEEbiography}

\begin{IEEEbiography}[{\includegraphics[width=1in,height=1.25in,clip,keepaspectratio]{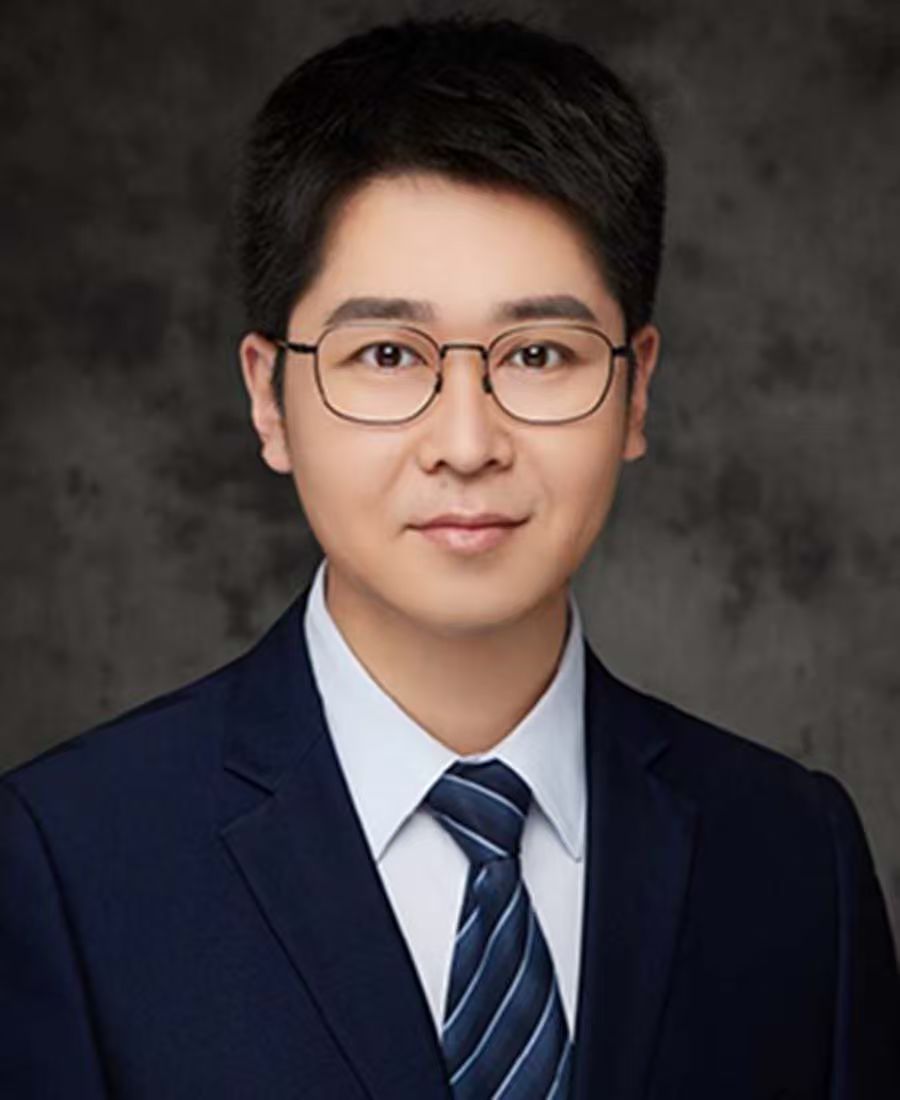}}]{Peng Shi}
received the Ph.D. degree in aerospace vehicle design from Beihang University, Beijing, China, in 2009. 
He is currently a Professor with the School of Astronautics and holds the position of Vice Dean at the institute. His research interests include satellite formation flight dynamics and control, spacecraft orbit and attitude dynamics, spacecraft guidance, navigation, and control, spacecraft intelligent game system, advanced control theory, and its applications in the aerospace domain.
\end{IEEEbiography}

\begin{IEEEbiography}[{\includegraphics[width=1in,height=1.25in,clip,keepaspectratio]{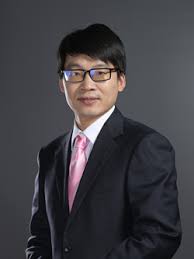}}]{Shengping Gong} received his B.S. degree in aerospace engineering from the National University of Defense Technology, Changsha, China, in 2004, and the Ph.D. degree in astrodynamics and control from Tsinghua University, Beijing, China, in 2008.

After spending a year as a Postdoctoral Researcher, he became an Assistant Researcher and then an Associate Professor with the School of Aerospace Engineering, Tsinghua University. In 2021, he was a Professor with the School of Astronautics, Beihang University. His research interests include the dynamics and control of spacecraft, trajectory optimization, solar sail, and celestial mechanics.
\end{IEEEbiography}

\end{document}